\documentclass[preprint]{revtex4}
\def\varepsilon{\epsilon}
\def\rm{\bf}
\def\iota{\imath}
\usepackage[psamsfonts]{amssymb}
\usepackage{amsmath}

\begin{document}

\title{Finite temperature Cherenkov radiation in the presence of a magnetodielectric medium}

\author{Fardin Kheirandish}
\email[]{fardin_kh@phys.ui.ac.ir}
\affiliation{Department of
Physics, Faculty of Science, University of Isfahan, Isfahan Iran.}
\collaboration{Quantum optics research group, University of
Isfahan, Isfahan, Iran.} \noaffiliation
\author{Ehsan Amooghorban}
\email[]{amooghorban@sci.ui.ac.ir}
\date{\today}
\begin{abstract}
A canonical approach to Cherenkov radiation in the presence of a magnetodielectric medium is presented in
classical, nonrelativistic and relativistic quantum regimes. The equations of motion for the canonical variables
are solved explicitly for both positive and negative times. Maxwell and related constitute equations are
obtained. In the large-time limit, the vector potential operator is found and expressed in terms of the medium
operators. The energy loss of a charged particle, emitted in the form of radiation, in finite temperature is
calculated. A Dirac equation concerning the relativistic motion of the particle in presence of the
magnetodielectric medium is derived and the relativistic Cherenkov radiation at zero and finite temperature is
investigated. Finally, it is shown that the Cherenkov radiation in nonrelativistic and relativistic quantum
regimes, unlike its classical counterpart, introduces automatically a cutoff for higher frequencies beyond which
the power of radiation emission is zero.
\end{abstract}

\maketitle
\section{Introduction}
Cherenkov radiation is the radiation with continuous spectrum and specific angular distribution emitted by the
medium due to the motion of charged particles moving in the medium with a velocity exceeding the phase velocity
of light in the transparent medium. The Cherenkov radiation in transparent media was experimentally observed by
Cherenkov \cite{1}. Theoretical explanation on this phenomenon was first developed by Frank and Tamm \cite{2}.
They showed that the particle should radiate when its velocity exceeds the velocity of light in the medium as
the emitted rays make an angle $\theta$ with the charge velocity given by $\cos\theta={c \mathord{\left/
 {\vphantom {v {n}}} \right.
 \kern-\nulldelimiterspace} {vn}}$ where $v$ is the speed of the
particle, $n$ is the index of refraction of the medium. Also they are polarized with the electric field in the
plane of this angle. The radiation chock-front, called the Cherenkov cone, is analogous to the Mach cone formed
as objects move with supersonic speeds through air. The cherenkov theory has drawn a great deal of attention all
around the world. This theory has been widely used in high-energy particle physics, optics, cosmic-ray physics,
high-power radiation sources, and so on \cite{3}-\cite{6}. Typical examples are the discoveries of the
anti-proton \cite{7} and the J-particle \cite{8}.

The classical theory of Cherenkov effect is sufficiently accurate in
the optical part of spectrum. For methodological reasons, it is
equally important to consider the quantum theory of this effects.
The phenomenological quantum theory of the Cherenkov radiation
developed by Ginzburg but the dissipative character of the medium
was neglected \cite{9}-\cite{10}. The source-theory explanation of
this effect was given by Schwinger et al. \cite{11}. Unfortunately,
due to the method of combining the denominators of the propagators
in parametric form, the resulting integrals are exceedingly
complicated and approximations were necessarily made.

The first formulation of the finite-temperature quantum field theory
was presented by Dolan and Jackiw \cite{12}, Weinberg \cite{13} and
Bernard \cite{14} and the first application of it concerned the
effective potentials in Higgs theories. Also, the inclusion of
temperature has been carried out in QED of Cherenkov radiation only
in a nondispersive dielectric medium \cite{15}. The main purpose of
the present work is to develop a canonical theory of the finite
temperature Cherenkov radiation to evaluate the electromagnetic
field arising from the uniform motion of a charged particle in the
presence of a magnetodielectric medium in different regimes i.e,
classical, non-relativistic and relativistic quantum regimes. To
achieve this goal we first generalize a Lagrangian introduced in
\cite {16} to include the external charges. This prepares not only
the grounds to extend the Ginzburg theory to include the dissipative
character and the permeable character of a medium but extracts a
Dirac equation for a relativistic particle in the presence of the
magnetodielectric medium to survey the relativistic effects of the
Cherenkov radiation to finite temperature regime.

The layout of the paper is as follows: In section 2, a Lagrangian for the total system is proposed and a
classical treatment of the Cherenkov radiation to finite temperature regime is investigated. In sections 3, we
use the Lagrangian introduced in the section 2 to canonically quantize the electromagnetic filed arises by the
moving external charges embedded in the magnetodielectric medium. Maxwell and constitute equations are obtained.
We find that for sufficiently large times the vector potential operator can be expressed in terms of the initial
medium operators. The consistency of these solutions for the vector potential operator depend on the validity of
certain velocity sum rules. We also show how to relate the results to the damping polarization model and
phenomenological quantization theories. By considering finite temperature effects, energy loss of a charged
particle emitted in the form of radiation is calculated. Subsequently, this formalism is generalized somehow to
describe the relativistic moving particles. Finally, we discuss the main results and conclude in section 3.

\section{Classical Theory}
Cherenkov radiation has the property that it occurs with uniform
motion of a charged particle in a spatially homogeneous medium
\cite{17}. At first we attempt to treat the theory of Cherenkov
radiation in the presence of a linear homogeneous magnetodielectric
medium on basis of the classical theory. In the first part of this
section we generalize the approach presented in \cite{16} to the
case where there are some external charges in the medium and in the
second part we examine the emission of electromagnetic wave, if it
occurs, by the moving particle. In the third section we will
concentrate our attention on the Cherenkov radiation in the finite
temperature situation.
\subsection{Classical dynamics}
Classical and quantum electrodynamics in a linear magnetodielectric,
can be accomplished by modeling the medium with two independent
reservoirs that interact with electromagnetic field. Each reservoir
contains a continuum of three dimensional harmonic oscillators
describing the polarizability and magnetizability properties of the
medium \cite{18}. Therefore, in order to have a classical treatment
of electrodynamics in a homogenous magnetodielectric medium, we
begin with the following classical Lagrangian for the total system
(medium+electromagnetic field+external charges)
\begin{equation} \label{a1}
L(t) = L_{res}  + L_{em} +L_{q}+L_{int}.
\end{equation}
The first term $L_{res}$ is the reservoir part
\begin{eqnarray} \label{a2}
 L_{res} \,& =& \int_0^\infty  {d\omega \int {d^3{\rm r}\,
 \frac{1}{2}[{\rm \dot X}_\omega({\rm r},t)   \cdot {\rm \dot X}_\omega({\rm r},t)-
\omega ^2 {\rm X}_\omega({\rm r},t)   \cdot {\rm X}_\omega({\rm r},t) ]}}\\ \nonumber &+&  \int_0^\infty
{d\omega \int {d^3{\rm r}\, \frac{1}{2}[{\rm \dot Y}_\omega({\rm r},t)\cdot {\rm \dot Y}_\omega({\rm r},t) -
\omega ^2 {\rm Y}_\omega({\rm r},t) \cdot {\rm Y}_\omega({\rm r},t)]}}
\end{eqnarray}
where the dynamical variables ${\rm X}_\omega(\rm r,t)$ and ${\rm Y}_\omega(\rm r,t)$ correspond to the electric
and magnetic characters of the medium, respectively. The second term $L_{em}$ is the electromagnetic field
\begin{equation} \label{a3}
L_{em}  = \int {d^3{\rm r }[\frac{1}{2}\varepsilon _0 {\rm E}^2({\rm r},t)  - \frac{{{\rm B}^2({\rm r},t)
}}{{2\mu _0 }}]}.
\end{equation}
The third term $L_{q}$ is the Lagrangian of the external charges with mass $m_\alpha$ and position ${\rm
r}_\alpha$
\begin{equation}  \label{a4}
L_{q} \, = \frac{1}{2}\sum_\alpha{m_\alpha\rm \dot r}^2_\alpha(t) +\sum_\alpha (q_\alpha{\rm \dot r}_\alpha
\cdot {\rm A}({\rm
r}_\alpha,t)-q_\alpha\phi ({\rm r}_\alpha,t)),\\
\end{equation}
and finally $ L_{int} $ is the interaction term, which includes the linear interaction between the medium and
electromagnetic field trough coupling functions $f(\omega)$ and $g(\omega)$ and also the interaction between the
external charges and electromagnetic field
\begin{equation} \label{a5}
L_{int} = \int_0^\infty  {d\omega \int {d^3 {\rm r}} } \,\,f(\omega) {\bf X}_\omega  ({\rm r},t)\cdot {\bf
E}({\rm r},t) + \int_0^\infty  {d\omega \int {d^3 {\rm r}} } \,\,g(\omega) {\bf Y}_\omega  ({\rm r},t)\cdot {\bf
B}({\rm r},t)
\end{equation}
In the equations (\ref{a3}) and (\ref{a5}), ${\rm E} = -
\frac{{\partial {\rm A}}}{{\partial t}} - \nabla \phi$ and ${\rm B}
= \nabla \times {\rm A} $ are the total electric and magnetic fields
and $\rm A$ and $\phi$ are the vector and the scaler potentials. For
simplicity we work in the reciprocal space and write the fields in
terms of their spatial Fourier transforms. The range of the variable
${\rm k}$ in the reciprocal space is restricted to the half space
denoted by a prime over the integral i.e., $\int^{'}d^3{\rm k}$
\cite{19}, thus in the reciprocal half space the Lagrangian
(\ref{a1}) can be written as
\begin{eqnarray} \label{a6}
\underline L(t)=\underline{L} _{q}(t)+\underline L _{res} (t)
+\underline L _{em} (t)+\underline L _{ int } (t)
\end{eqnarray}
\begin{equation}\label{a7}
\underline{L}_{q} \, = \frac{1}{2}\sum_\alpha{m_\alpha\rm \dot
r}^2_\alpha(t)+ \frac{1}{{(2\pi )^{{3 \mathord{\left/ {\vphantom {3
2}} \right. \kern-\nulldelimiterspace} 2}} }}\sum_\alpha\int^{'}
{d^3 {\rm k}} \left[ q_\alpha{\left( {{\rm \dot r} _\alpha\cdot
\underline {\rm A} ({\rm k},t) - \underline \phi  ({\rm k},t)}
\right)e^{\iota {\rm k} \cdot {\rm r}_\alpha} + c.c.} \right],
\end{equation}
\begin{equation}\label{a8}
\underline L _{res} (t) = \int_0^\infty  {d\omega \int^{'} {d^3{\rm k}(|{\underline {{\rm \dot X}} _\omega }|^2
- \omega ^2 |{\underline {\rm X} _\omega }|^2 )} }+\int_0^\infty  {d\omega \int^{'} {d^3{\rm k}(| {\underline
{{\rm \dot Y}} _\omega  }|^2 - \omega ^2 | {\underline {\rm Y} _\omega  }|^2 )} }
\end{equation}
\begin{equation}\label{a9}
\underline L _{em} (t) = \int^{'} {d^3{\rm k}(\varepsilon _0 |{\underline {{\rm \dot A}} }|^2  + \varepsilon _0
|{{\rm k}\underline \phi }|^2  - \frac{{|{{\rm k} \times \underline {\rm A} }|^2 }}{{\mu _0 }})} + \varepsilon
_0\int^{'} {d^3{\rm k}} ( - \iota {\rm k} \cdot \underline {{\rm \dot A}} \underline \phi ^{\rm *} +
 c.c.)
\end{equation}
\begin{eqnarray}\label{a10}
\underline L _{int} & =&- \int_0^\infty  {d\omega \int^{'} {d^3 {\rm k}} } \,\,\left[ { f (\omega)\, \underline
{\rm X}^*_\omega  ({\rm k},t) \cdot(\imath{\bf k}\underline{\varphi}({\rm k},t)+\underline{\dot{{\bf A}}}({\rm
k},t))+ c.c.} \right]
\nonumber\\
&+&\int_0^\infty  {d\omega \int^{'} {d^3 {\rm k}} } \,\,\left[ { g (\omega)\, \underline {\rm Y}^*_\omega  ({\rm
k},t) \cdot(\imath{\bf k}\times\underline{{\bf A}}({\rm k},t))+ c.c.} \right]
\end{eqnarray}
where we have applied $\underline {\rm X}^* ({\rm k},t)=\underline
{\rm X} ({\rm -k},t)$ and the similar relations for the other
dynamical fields \cite{19}. we can obtain the classical equations of
the motion simply from Euler-Lagrange equations. For the vector and
scalar potentials ${\rm A} ({\rm k},t)$, $\phi({\rm k},t)$ we have
\begin{eqnarray} \label{a11}
&&\frac{d}{{dt}}\left( {\frac{{\delta \underline L }}{{\delta (\underline {{\rm \dot A}} _i^* ({\rm k},t))}}}
\right) - \frac{{\delta \underline L }}{{\delta (\underline {\rm A} _i^*
({\rm k},t))}} = 0 \hspace{+1cm}i=1,2,3 \nonumber\\
&& \Rightarrow \mu _0 \varepsilon _0 \underline {{\rm \ddot A}} ({\rm k},t) + \mu _0 \varepsilon _0 \iota {\rm
k}\underline {\dot \phi } ({\rm k},t) - {\rm k} \times \left( {{\rm k} \times
\underline {\rm A} ({\rm k},t)} \right) =\nonumber\\
&&\hspace{1cm}\mu _0\underline{\dot{{\bf P}}}({\rm k},t)+\imath\mu _0 {\bf k }\times\underline{{\bf M }}({\rm
k},t)+\mu _0 \underline {\rm J} ({\rm k},t)
\end{eqnarray}
and
\begin{eqnarray} \label{a12}
&& \frac{d}{{dt}}\left( {\frac{{\delta \underline L }}{{\delta (\underline {\dot \phi } ^* ({\rm k},t))}}}
\right) - \frac{{\delta \underline L }}{{\delta (\underline \phi  ^* ({\rm
k},t))}} = 0 \nonumber\\
 && \Rightarrow- \varepsilon _0 \iota {\rm k} \cdot
\underline{{\rm \dot A}}({\rm k},t) + \varepsilon _0 {\rm k}^2
\underline \phi ({\rm k},t) =-\imath{\bf k }\cdot\underline{{\bf
P}}({\rm k},t)+ \underline \rho  ({\rm k},t)
\end{eqnarray}
where
\begin{eqnarray} \label{a13}
&&\underline{{\rm P}}({\rm k},t) = \int_0^\infty  {d\omega } \,f(\omega )\,{\rm
\underline{X}}_\omega  ({\rm k},t),\\
&&\underline{{\rm M}}({\rm k},t) = \int_0^\infty  {d\omega } \,g(\omega )\,{\rm \underline{Y}}_\omega  ({\rm
k},t),
\end{eqnarray}
\begin{eqnarray}\label{a13b}
&& \underline {\rm J} ({\rm k},t) = \frac{1}{{(2\pi )^{{3 \mathord{\left/ {\vphantom {3 2}} \right.
\kern-\nulldelimiterspace} 2}} }}\sum_\alpha\,\,\,q_\alpha{\rm \dot r}_\alpha\,\,\,
e^{ - \iota {\rm k} \cdot {\rm r}_\alpha},  \\
&&\underline \rho  ({\rm k},t) = \frac{1}{{(2\pi )^{{3 \mathord{\left/ {\vphantom {3 2}} \right.
\kern-\nulldelimiterspace} 2}} }}\sum_\alpha q_\alpha e^{ - \iota {\rm k} \cdot {\rm r}_\alpha},
\end{eqnarray}
are respectively the Fourier transforms of the electric and magnetic
polarization densities of the medium. The Fourier transforms of the
external current and charge densities satisfy the charge
conservation in the reciprocal space
$\underline{\dot{\rho}}=-\imath{\bf k}\cdot \underline{{\bf J}}$.
Similarly from the Euler-Lagrange equations for the fields ${\bf
r}_\alpha$ and $\underline{\rm X}_{\omega}$, $\underline{\rm
Y}_{\omega}$ we find
\begin{eqnarray}\label{a14}
&& \frac{d}{{dt}}\left( {\frac{{\delta \underline L }}{{\delta (\dot r_{\alpha,i} (t))}}} \right) -
\frac{{\delta \underline L }}{{\delta
(r_{\alpha,i} (t))}} = 0\hspace{+2cm}i=1,2,3  \nonumber\\
&&\Rightarrow m_\alpha{\rm \ddot r}_\alpha(t)= q_\alpha{\rm E}({\rm r}_\alpha,t) + q_\alpha\dot {\rm {r}_\alpha}
\times {\rm B}({\rm r}_\alpha,t)
\end{eqnarray}
and
\begin{eqnarray}\label{a15}
&& \frac{d}{{dt}}\left( {\frac{{\delta \underline L }}{{\delta (\underline {{\rm \dot X}} _{\omega i}^* ({\rm
k},t))}}} \right) - \frac{{\delta \underline L }}{{\delta (\underline {\rm X} _{\omega
i}^* ({\rm k},t))}} = 0 \hspace{+2cm}i=1,2,3 \nonumber\\\nonumber\\
&& \hspace{-1cm}\Rightarrow \underline {{\rm \ddot X}} _{\omega }
({\rm k},t) + \omega ^2 \underline {\rm X} _{\omega } ({\rm k},t) =
- f(\omega )\,(\underline{{\rm \dot A}}({\rm k},t)\, + \iota
k\underline{\varphi} ({\rm k},t))
\end{eqnarray}
\begin{eqnarray}\label{a16}
&& \frac{d}{{dt}}\left( {\frac{{\delta \underline L }}{{\delta (\underline {{\rm \dot Y}} _{\omega i}^* ({\rm
k},t))}}} \right) - \frac{{\delta \underline L }}{{\delta (\underline {\rm Y} _{\omega
i}^* ({\rm k},t))}} = 0 \hspace{+2cm}i=1,2,3 \nonumber\\\nonumber\\
&& \hspace{-1cm}\Rightarrow \underline {{\rm \ddot Y}} _{\omega}
({\rm k},t) + \omega ^2 \underline {\rm Y} _{\omega } ({\rm k},t) =
g(\omega )\,\imath{\rm k}\times(\underline{{\rm \dot A}}({\rm k},t).
\end{eqnarray}
The formal solution of the field equation (\ref{a15}) is
\begin{equation} \label{a17}
\underline {\rm X}_\omega ({\rm k},t) = \underline {{\rm \dot X}} _\omega ({\rm k},0)\frac{{\sin \omega
t}}{\omega } + \underline {\rm X} _\omega ({\rm k},0)\cos \omega t + f(\omega)\int_0^t {dt'} \frac{{\sin \omega
(t - t')}}{\omega } {\rm \underline{E}}({\rm k},t'),
\end{equation}
where the first term is the inhomogeneous solution of the equation
(\ref{a15}) and the second term is the homogeneous one. We will show
that after quantization the homogeneous solution becomes a noise
operator. However, since we are only interested in the induced
polarization, we keep the inhomogeneous solution and using the
equation (\ref{a13}), we find the electric polarization density of
the medium in reciprocal space
\begin{equation}\label{a18}
\underline{{\rm P}}({\rm k},t) = \epsilon_0\int_0^\infty  {dt\,} \chi_e(t - t'){\bf \underline{E}}({\rm k},t')
\end{equation}
where $\chi_e$ is the electric causal susceptibility of the medium and in terms of the coupling function $f$ can
be written as
\begin{eqnarray} \label{a19}
\chi_e(t-t') =\left\{ \begin{array}{l} \frac{1}{\epsilon_0}\int_0^\infty  {d\omega } \frac{{\sin \omega
(t-t')}}{\omega } f^2 (\omega )\hspace{2cm}t > t' \\
\\
0\hspace{5cm}t < t' \\
\end{array} \right.
\end{eqnarray}
which is the origin of the significant Kramers-Kronig relations
\cite{17}. In a similar fashion the magnetic polarization density of
the medium in reciprocal space can be obtained straightforwardly
using the formal solution of the equation (\ref{a16}) as
\begin{equation}\label{a22}
\underline{{\rm M}}({\rm k},t) = \frac{1}{\mu_0}\int_0^\infty {dt\,} \chi_m(t - t'){\bf \underline{B}}({\rm
k},t')
\end{equation}
where $\chi_m$ is the magnetic causal susceptibility of the medium which in terms of the coupling function $g$
can be written as
\begin{eqnarray} \label{a23}
\chi_m(t-t') =\left\{ \begin{array}{l}
 \mu_0\int_0^\infty  {d\omega } \frac{{\sin \omega (t-t')}}{\omega } g^2 (\omega )\hspace{2cm}t > t' \\
\\
0\hspace{5.1cm}t < t'\\
\end{array} \right.
\end{eqnarray}
The electric permittivity and the inverse magnetic permeability of the magnetodielectric medium are defined in
terms of $\chi_e$ and $\chi_m$ as
\begin{equation}\label{a24}
\epsilon(\omega)=1+\chi_e(\omega)
\end{equation}
and
\begin{equation}\label{a25}
\kappa(\omega)=1-\chi_m(\omega)
\end{equation}
where
\begin{equation}\label{a26}
\chi_{e,m} (\omega) = \int_{0}^\infty {dt }\chi_{e,m}(t)
e^{\imath\omega t}\nonumber
\end{equation}
By using Eqs. (\ref{a19}) and (\ref{a23}), we can obtain the
following important relations in the frequency domain
\begin{eqnarray}\label{a26/1}
\chi_{e} (\omega) &=&\frac{1}{\epsilon_0} \int_0^\infty {d\omega'
}\frac{f^2 (\omega'  )}{\omega'^2-\omega^2-\imath0^+}
\end{eqnarray}
\begin{equation}\label{a26/2}
\chi_{m} (\omega) ={\mu_0} \int_0^\infty {d\omega' }\frac{g^2
(\omega'  )}{\omega'^2-\omega^2-\imath0^+}.
\end{equation}
These are complex functions of frequency which satisfy
Kramers-Kronig relations and have the properties of response
functions i.e, $\epsilon(-w^*)=\epsilon^*(\omega)$ and
$\kappa(-w^*)=\kappa^*(\omega)$ and $Im \epsilon(\omega)>0$, $Im
\mu(\omega)>0$ provided that $f^2(-\omega^*)=f^2(\omega)$ and
$g^2(-\omega^*)=g^2(\omega)$. It can be shown that these functions
have no poles in the upper half plane and tend to zero as
$\omega\longrightarrow\infty$. If we are given definite electric
permittivity and inverse magnetic permeability of the medium then we
can inverse the relations (\ref{a19}) and (\ref{a23}) and find the
corresponding coupling functions $f(\omega)$ and $g(\omega)$ as
\begin{eqnarray}\label{a27}
f(\omega ) = \sqrt{\frac{{2\omega
\epsilon_0}}{\pi}Im\epsilon(\omega)}
\end{eqnarray}
\begin{equation}\label{a27/0}
g(\omega ) = \sqrt{-\frac{{2\omega }}{\pi\mu_0}Im\kappa(\omega)}
\end{equation}

where the minus sign in the second expression is used since for a magnetodielectric medium $Im \mu(\omega)>0$,
therefore $Im \kappa(\omega)<0$. In order to illustrate the relations between the coupling functions and the
electric permittivity and the inverse magnetic permeability of a magnetodielectric medium, let us restrict our
attention to a single resonance electric permittivity which can be obtained from the Lorentz oscillator model
\begin{equation}\label{a27/1}
\epsilon(\omega)=1+\frac{\omega_{pe}^2}{\omega_{0e}^2-\omega^2-\imath\gamma_e\omega}
\end{equation}
and a single resonance magnetic permeability \cite{20}
\begin{equation}\label{a27/2}
\mu(\omega)=1+\frac{\omega_{pm}^2}{\omega_{0m}^2-\omega^2-\imath\gamma_m\omega}
\end{equation}
where $\omega_{pe}$ and $\omega_{pm}$ are the coupling strengths,
$\omega_{0m}$, $\omega_{0m}$ are the transverse resonance
frequencies, and $\gamma_e$, $\gamma_m$ are the absorption
parameters. Now by using Eqs. (\ref{a27})  and (\ref{a27/0}) the
coupling functions $f(\omega)$ and $g(\omega)$ can be obtained as
\begin{equation}\label{a27/2}
f^2(\omega)=\frac{2\gamma_e\epsilon_0\omega_{pe}^2\omega^2/\pi}{(\omega_{0e}^2-\omega^2)^2+\gamma_e^2\omega^2},
\end{equation}
\begin{equation}\label{a27/3}
g^2(\omega)=\frac{2\gamma_m\omega_{pm}^2\omega^2/\pi\mu_0}{(\omega_{0m}^2+\omega_{pm}^2-\omega^2)^2+\gamma_m^2\omega^2}.
\end{equation}

Following the standard approach to classical electrodynamics, we
choose the Coulomb gauge ${\bf k}\cdot{\bf A}=0$, so that the vector
potential ${\bf A}$ is a purely transverse field. By useing the
Euler-Lagrange equation for $\dot{\varphi}^*$, we eliminate
$\varphi$ from equation (\ref{a11}) and substituting Eqs.
(\ref{a18}) and (\ref{a22}) into Eq. (\ref{a11}), obtain the
following inhomogeneous wave equation
\begin{eqnarray}\label{a28}
&&\mu _0 \varepsilon _0 \underline{{\rm \ddot A}}({\rm k},t) + k^2 \underline{{\rm A}}({\rm k},t) - \mu _0 k^2
\int_0^t {dt'} \chi _m (t - t')\,\underline{{\rm A}}({\rm k},t')+ \mu _0 \frac{\partial }{{\partial t}}\int_0^t
{dt'} \chi _e (t - t')\,\underline{{\rm \dot A}}({\rm k},t')\nonumber\\
 &&= \mu _0 \underline{{\rm J}}^ \bot ({\rm
k},t).
\end{eqnarray}
where the transverse current is defined as ${\underline{{\bf J}}^\bot}=\sum_{\lambda=1}^2\underline{{\bf
J}}\cdot e_\lambda({\bf k}) $ with unit polarization vectors $e_\lambda({\bf k}), \lambda=1,2$, which are
orthogonal to $e_3 ({\rm k}) =\frac{\bf k}{k}= \hat k $ and to one another. This equation can be solved in terms
of initial conditions using the Laplace transforms. For any time-dependent operator $\Omega(t)$ the forward
Laplace transform is defined as
\begin{eqnarray}\label{a29}
\tilde{\Omega}^f (s) = \int_0^\infty  {dt} e^{ - st} \Omega (t),
\end{eqnarray}
obviously, the Laplace transform contains all information about the time evolution of $\Omega$ for positive $t$.
In the following we wish to determine the time evolution of the relevant operators of our model for any time,
either positive or negative. Hence, we also introduce the backward Laplace transform
\begin{eqnarray}\label{a30}
\tilde{\Omega}^b (s) = \int_0^\infty  {dt} e^{ - st}\Omega (-t).
\end{eqnarray}
Let $\tilde{\epsilon}(s)$ and $\tilde{\kappa}(s)$ be the Laplace
transformations of $\epsilon(t)$ and $\kappa(t)$, respectively. Then
$\underline{\tilde{A}}^f({\bf k},s)$ and
$\underline{\tilde{A}}^b({\bf k},s)$ i.e., the forward and backward
Laplace transformation of $\underline{A}({\bf k},t)$, can be
obtained as follows
\begin{eqnarray}\label{a31}
\tilde{\underline{{\rm  A}}}^{f,b} ({\rm k},s) &=
&\frac{{s\tilde{\epsilon }(s)}}{{s^2 \tilde{\epsilon} (s) + k^2 c^2
\tilde{\kappa} (s)}}\underline{{{\rm A}}}({\rm k},0) \pm
\frac{1}{{s^2 \tilde{\epsilon} (s) + k^2 c^2\tilde{ \kappa}
(s)}}\underline{{\dot
{{\rm A}}}}({\rm k},0)\nonumber\\
 &+& \frac{1}{{\epsilon _0
}}\sum_\lambda\frac{{(\underline{\tilde{{\rm  J}}}^{f,b} ({\rm k},s)
\cdot e_\lambda  ({\rm k}))e_\lambda ({\rm k})}}{{s^2
\tilde{\epsilon} (s) + k^2 c^2 \tilde{\kappa }(s)}}
\end{eqnarray}
The time-dependent vector potential is obtained from a contour
integration over the Bromwich contour by an inverse Laplace
transformation. From Eq. (\ref{a31}) we find the vector potential
for $t>0$, \cite{21}
\begin{eqnarray}\label{a32}
\underline{{\rm A}}({\rm k},t) &=& \,\xi(t)\underline{A}({\rm k},0) + \zeta(t)\underline{\dot A}({\rm k},0)\nonumber\\
& -& \frac{1}{{2\pi \varepsilon _0 }}\sum_\lambda\int_{ - \infty }^{
+ \infty }d\omega e^{ - \iota \omega t}
{\frac{{{(\underline{\tilde{{\rm J}}}^f({\rm k},-\imath\omega+0)
\cdot e_\lambda  ({\rm k}))} }}{\omega^2 \epsilon (\omega) - k^2 c^2
\kappa (\omega)}}e_\lambda ({\rm k}),
\end{eqnarray}
where
\begin{equation}\label{a33}
\xi (t) = \frac{1}{{2\pi \iota }}\int_{ - \iota \infty }^{ + \iota
\infty } dse^{st}{\frac{{s\tilde{\epsilon}(s) }}{s^2
\tilde{\epsilon} (s) + k^2 c^2 \tilde{\kappa} (s)}}= \sum\limits_j
{} {\mathop{Re}\nolimits} (e^{ - \iota \Omega _j t} \frac{{v_g^j
}}{{v_p^j }})
\end{equation}
and
\begin{equation}\label{a34}
\zeta (t) = \frac{1}{{2\pi \iota }}\int_{ - \iota \infty }^{ + \iota
\infty } ds{\frac{{e^{st} }}{s^2\tilde{ \epsilon} (s) + k^2 c^2
\tilde{\kappa} (s)}}=\frac{1}{{kc}}\sum\limits_j  {\mathop{
Im}\nolimits} (e^{ - \iota \Omega _j t} \frac{{v_g^j }}{{c{\kappa}
(\Omega _j )}})
\end{equation}
In these expressions, we changed the integration variable from $s$
to $-\imath\omega+\eta$, with a small but positive $\eta$.
Therefore, we introduce the electric permittivity and the inverse
magnetic permeability of the magnetodielectric medium in the
frequency domain as
$\epsilon(\omega)=\tilde{\epsilon}(-\imath\omega+0)$ and
$\kappa(\omega)=\tilde{\kappa}(-\imath\omega+0)$ for real $\omega$,
and correspondingly, $\tilde{\epsilon}(\imath\omega+0)$ and
$\tilde{\kappa}(\imath\omega+0)$ as their complex conjugations
$\epsilon^*(\omega)$ and $\kappa^*(\omega)$, respectively. We define
for each allowed frequency $\Omega_j$, the group velocity
$v_g^j=\frac{\partial{\omega}}{\partial k}$ and the phase velocity
$v_p^j=\frac{\omega}{k}$ where the frequencies $\Omega_j({\bf k})$
and $\Omega_j^*({\bf k})$ are the complex-frequency solutions of the
dispersion relation $\omega^2\epsilon(\omega)-{\bf
k}^2c^2\kappa(\omega)$ which has no zeros in the upper half plane.
It is worth emphasizing that for a lossy medium, we lose the usual
dispersion relation in which a limited number of discrete
frequencies $\omega$ are associated with each wave vector $k$. Thus
$k$ and $\omega$ must be considered as independent real variables
\cite{22}.

Now the vector potential for $t<0$ is obtained from the inverse Laplace transform of Eq. (\ref{a31}) as
\begin{eqnarray}\label{a35}
\underline{{\rm A}}({\rm k},t) &=& \,\xi(t)\underline{{\rm A}}({\rm k},0) - \zeta(t)\underline{\dot {\rm A}}
({\rm k},0)\nonumber\\
& -& \frac{1}{{2\pi\varepsilon _0 }}\sum_\lambda\int_{ - \infty }^{
+ \infty }d\omega e^{ - \iota \omega t}
{\frac{{{(\underline{\tilde{{\rm J}}}^b({\rm k},+\imath\omega+0)
\cdot e_\lambda  ({\rm k}))} }}{\omega^2 \epsilon^* (\omega) - k^2
c^2 \kappa ^*(\omega)}}e_\lambda ({\rm k}),
\end{eqnarray}
The coefficients $\eta(t)$ and $\zeta(t)$ in Eqs. (\ref{a32}) and (\ref{a35}) damp out exponentially in time
since all $\Omega_j$ in the exponentials have negative imaginary parts. Therefore, For large times, the medium
and electromagnetic field tend to an equilibrium which is determined by the characteristic damping time
$\tau_j={1 \mathord{\left/
 {\vphantom {1 {{\mathop{\rm Im}\nolimits} \Omega _j }}} \right.
 \kern-\nulldelimiterspace} {{\mathop{ Im}\nolimits} \Omega _j }}
$. After a few times the maximum characteristic damping time, only
the third term survives in these equations since they have poles on
the imaginary axis in the complex $s$ plane. Also, these terms in
Eqs. (\ref{a32}) and (\ref{a35}) are zero for negative and positive
$t$, respectively \cite{23}. Thus, we may combine the two
expressions into a single one and use the Fourier transform to
obtain the vector potential in real space for all $t$
\begin{equation}\label{a36}
{\rm A}({\rm r},t) = \int_{0 }^{ + \infty }d\omega\int {d^3 {\rm k}}
\,e^{\imath{\bf k}\cdot{\bf r} - \iota \omega t}\underline{{\rm
A}}^+({\rm k},\omega)+c.c.,
\end{equation}
with the positive-frequency Fourier component
\begin{equation}\label{a37}
\underline{{\rm A}}^+({\rm k},\omega) = \frac{-1}{{(2\pi )^{{5
\mathord{\left/ {\vphantom {5 2}} \right. \kern-\nulldelimiterspace}
2}} \varepsilon _0 }}
\sum_\lambda\big[{\frac{{{\underline{{\tilde{{\rm J}}}}^f({\rm
k},-\imath\omega+0) \cdot e_\lambda ({\rm k})} }}{\omega^2 \epsilon
(\omega) - k^2 c^2 \kappa (\omega)}}+
{\frac{{{\underline{{\tilde{{\rm J}}}}^b({\rm k},\imath\omega+0)
\cdot e_\lambda ({\rm k})} }}{\omega^2 \epsilon^* (\omega) - k^2 c^2
\kappa ^*(\omega)}}\big]e_\lambda ({\rm k})
\end{equation}
It is not difficult to show that the transverse electric field can be written as
\begin{eqnarray}\label{a38}
{\rm E}^\bot({\rm r},t) = -\imath\int_{0 }^{ + \infty
}d\omega\,\omega\int {d^3 {\rm k}} \,(e^{\imath{\bf k}\cdot{\bf r} -
\iota \omega t}\underline{{\rm A}}^{+}({\rm k},\omega)-c.c.)
\end{eqnarray}
From the Eq. (\ref{a32}) we can see that the coefficient $\xi(t)$ in Eq. (\ref{a33}) takes the value 1 at time
$t=0$ and therefore the coefficient $\zeta(t)$ takes the value 0. These constraints are satisfied, if certain
velocity sum rules are adopted. In this way, we find the following modified velocity sum rules for all wave
vectors $\bf k$
\begin{equation}\label{a39}
\sum\limits_j  {\mathop{Re}\nolimits} ( \frac{{v_g^j }}{{v_p^j }})=1
\end{equation}
and
\begin{equation}\label{a40}
\sum\limits_j  {\mathop{ Im}\nolimits} ( \frac{{v_g^j }}{{\kappa
(\Omega _j )}})=0,
\end{equation}
which are resembling the quantum relations obtained in \cite{21} and
\cite{24}, now generalized to a magnetodielectric medium.
\subsection{Classical theory of Cherenkov radiation (T=0)}
Theoretically, when considering the Cherenkov radiation, one usually treats the charge motion with a constant
velocity which corresponds to the so-called Tamm– Frank problem \cite{2}. Consider a point charge $e$ uniformly
moving in a magnetodielectric medium with a velocity $\bf v$. Therefore, according to Eq. (\ref{a13b})
\begin{eqnarray}\label{a41}
{\rm \underline{J}}({\rm k},t) = \frac{e{\rm v}}{{(2\pi )^{{3 \mathord{\left/ {\vphantom {3 2}} \right.
\kern-\nulldelimiterspace} 2}} }}e^{ - \iota {\rm k} \cdot {\rm v}t},
\end{eqnarray}
then
\begin{equation}\label{a42}
{\underline{\tilde{{\rm J}}}^f}({\rm k},-\imath\omega+0)=\frac{e{\rm
v}}{{(2\pi )^{{1 \mathord{\left/ {\vphantom {3 2}} \right.
\kern-\nulldelimiterspace} 2}} }}\delta(\omega-{\bf k}\cdot{\rm v}).
\end{equation}
Now substituting Eq. (\ref{a42}) into the equation (\ref{a38}), we obtain
\begin{equation}\label{a43}
{\rm E}^\bot({\rm r},t) =\frac{-\imath e}{8\pi^3\epsilon_0 }
\sum_\lambda\int_{0 }^{ + \infty }d\omega\,\omega\int {d^3 {\rm k}}
\,[ {\frac{{e^{\imath{\bf k}\cdot{\bf r}}e^{ - \iota \omega t}{{\bf
v} \cdot e_\lambda ({\rm k})} }}{\omega^2 \epsilon (\omega) - k^2
c^2 \kappa (\omega)}}-c.c ]\delta(\omega-{\bf k}\cdot{\rm
v})e_\lambda ({\rm k}).
\end{equation}
In this case the energy loss of a point charged particle per unit
length emitted in the form of radiation, is defined by the braking
force acting on the charge at its location \cite {17}, \cite {25}
\begin{eqnarray}\label{a45}
\frac{{dW}}{{dt}} &=& e{\rm v} \cdot {\rm E}^ \bot  \left| {_{{\rm
r} = {\rm v}t} } \right. \nonumber \\
&=&\frac{\imath e^2}{8\pi^3\epsilon_0 } \sum_\lambda\int_{0 }^{ +
\infty }d\omega\,\omega\int {d^3 {\rm k}} \,[ {\frac{ e^{\imath({\bf
k}\cdot{\rm v}-\omega) t}{({{\bf v} \cdot e_\lambda ({\rm k}))^2}
}}{k^2 c^2 \kappa
(\omega)-\omega^2 \epsilon (\omega)}}-c.c ]\delta(\omega-{\bf k}\cdot{\rm v}).\nonumber\\
\end{eqnarray}
Letting $\theta$ be the angle between $\bf v$ and $\bf k$, then $\sum_\lambda({\bf v}\cdot e_\lambda({\bf
k}))^2={\bf v}^2(1-cos ^2\theta)$. Therefore we find
\begin{eqnarray}\label{a46}
\frac{{dW}}{{dt}} &=& \frac{\imath e^2v}{4\pi^2\epsilon_0 }\int_{0 }^{ + \infty }d\omega\omega\int_{0 }^{ +
\infty } dk k \,\int_{-1 }^{ + 1}d(cos \theta)[ {\frac{ e^{\imath (k v cos \theta- \omega )t}{{(1-cos ^2\theta)}
}}{k^2 c^2 \kappa (\omega)-\omega^2 \varepsilon (\omega)}}-c.c ]\delta(cos \theta-\frac{\omega}{kv})\nonumber\\
&=& \frac{ e^2v}{2\pi^2 \epsilon_0}\int_{0 }^{ + \infty }d\omega\omega\int_{0 }^{ + \infty } {dk}k \, {\frac{
{{Im(k^2 c^2 \kappa (\omega)-\omega^2 \varepsilon (\omega))} }}{|k^2 c^2 \kappa (\omega)-\omega^2 \varepsilon
(\omega)|^2}}(1-\frac{\omega^2}{k^2v^2})
\end{eqnarray}
that the electromagnetic waves are emitted at an angle to the path of the particle determined by
\begin{equation}\label{a46/1}
cos \theta=\frac{\omega}{kv}.
\end{equation}
The transparent magnetodielectric medium can be considered in principle as a limiting case of the lossy
dispersive medium. Actually, there are some ranges of frequencies over which the imaginary parts of the
permittivity and permeability of the magnetodielectric medium can be ignored. In a transparent magnetodielectric
medium the energy loss of a charge a only a result of radiation. In this case the imaginary parts of the
electric permittivity and the inverse magnetic permeability must tend to zero, thus
\begin{eqnarray}\label{a47}
\mathop {\lim }\limits_{{\mathop{ Im}\nolimits} \,\varepsilon (\omega ),\,\,{\mathop{ Im}\nolimits} \,\mu
(\omega ) \to 0} {\frac{ {{Im(k^2 c^2 \kappa (\omega)-\omega^2 \varepsilon (\omega))} }}{|k^2 c^2 \kappa
(\omega)-\omega^2 \varepsilon (\omega)|^2}}&=& \pi\mu(\omega)\delta(n^2(\omega)\omega^2-k^2c^2)
\nonumber\\
&=&\sum_j\frac{\pi \mu(\Omega_j)v_j^g}{2\Omega_jn(\Omega_j)c}\delta(\omega-\Omega_j)\nonumber\\
\end{eqnarray}
where $n^2(\omega)=\varepsilon (\omega)\mu(\omega)$ and the frequencies $\Omega_j({\bf k})$ are the
complex-frequency solutions of the dispersion relation $\omega^2\epsilon(\omega)-{ k}^2c^2\kappa(\omega)$. Now
Substituting Eq. (\ref{a47}) into Eq. (\ref{a46}) and performing the integration over $\omega$ and converting
the integration over $k$ to an integration over $\Omega_j$, we obtain
\begin{equation}\label{a48}
\frac{{dW}}{{dt}}=\frac{ e^2v}{4\pi\epsilon_0 c^2 }\sum_j\int_{0 }^{ + \infty
}d\Omega_j\,\Omega_j\,\mu(\Omega_j)(1-\frac{c^2}{v^2n^2(\Omega_j)})
\end{equation}

It is seen from the Dirac $\delta$-function in Eqs. (\ref{a47}) and (\ref{a46}) that radiation is possible only
if the inequality $v>{c \mathord{\left/
 {\vphantom {c {n(\Omega _j )}}} \right.
 \kern-\nulldelimiterspace} {n(\Omega _j )}}$ is satisfied. We define the Cherenkov
cone $cos \theta={c \mathord{\left/
 {\vphantom {v {n(\Omega _j )}}} \right.
 \kern-\nulldelimiterspace} {vn(\Omega _j )}}$ corresponding to any
frequency $\Omega_j$ for which $v>{c \mathord{\left/
 {\vphantom {c {n(\Omega _j )}}} \right.
\kern-\nulldelimiterspace} {n(\Omega _j )}}$, where $\theta$ is the
angle between the wave vector $\bf k$ of the radiated
electromagnetic wave and the velocity of the particle $\bf v$. It is
easy to show that when there is only one frequency for each $k$,
then the equation (\ref{a48}) tends to the result of \cite{9},
\cite{17} and \cite{26}.
\subsection{Finite temperature Cherenkov radiation in classical regime}
Our considerations so far have been applied to zero temperature. The
generalization of the formalism for this case is straightforward. It
is known that the medium and electromagnetic field are in the
thermal equilibrium in this regime. The inclusion of temperature may
be done in the usual manner \cite{27}-\cite{29}. The finite
temperature expression, as is well known, is found by replacing the
frequency integral by a sum over Matsubara frequencies according to
the transition
\begin{equation}\label{a49}
\hbar \int_0^\infty  {\frac{{d\xi }}{{2\pi }}} f(\iota \xi_l )\,\,\,
\to \,\,\,k_B T\sum\limits_{l = 0}^{\infty\,\,\,'}  {f(\iota \xi_l
),\,\,\,\,\,\,\,\,\,\,\,\,\,\,\,\,\,\,\,\,\,\,}  \xi_l  = {{2\pi k_B
Tl} \mathord{\left/
 {\vphantom {{2\pi k_B Tl} \hbar }} \right.
 \kern-\nulldelimiterspace} \hbar }
\end{equation}
where $T$ and $k_B$ are the temperature and Boltzmann constant and the prime on the summation mark denotes that
the zeroth term is given half weight as is conventional. The effect of finite temperature on the energy loss in
the form of Cherenkov radiation can be easily taken into account by using Eqs. (\ref{a49}) and (\ref{a46})
\begin{eqnarray}\label{a50}
\frac{{dW}}{{dt}} = \frac{ \imath e^2vk_B T}{\pi \epsilon_0
\hbar}\sum\limits_{l = 0}^{\infty\,\,\,'}\xi_l\int_{0 }^{ + \infty }
{dk}k \, {\frac{ {{Im(k^2 c^2 \kappa (\imath\xi_l)+\xi_l^2
\varepsilon (\imath\xi_l))} }}{|k^2 c^2 \kappa (\imath\xi_l)+\xi_l^2
\varepsilon (\imath\xi_l)|^2}}(1+\frac{\xi_l^2}{k^2v^2}).
\end{eqnarray}
We know that the function $\coth(\hbar\omega/2k_BT)$ has an infinite number of poles at $\omega_l=\imath\xi_l$
and elsewhere is analytic and bounded. This enables us to write Eq. (\ref{a50}) as
\begin{equation}\label{a51}
\frac{{dW}}{{dt}} = \frac{ e^2v}{2\pi^2 \epsilon_0}\int_{0 }^{ +
\infty }d\omega \,\omega \coth(\frac{\hbar\omega}{2k_BT})\int_{0 }^{
+ \infty } {dk}k \, {\frac{ {{Im(k^2 c^2 \kappa (\omega)-\omega^2
\varepsilon (\omega))} }}{|k^2 c^2 \kappa (\omega)-\omega^2
\varepsilon (\omega)|^2}}(1-\frac{\omega^2}{k^2v^2})
\end{equation}
Then with suitable rearrangement of the exponentials in the
hyperbolic cotangent we obtain
\begin{equation}\label{a52}
\frac{{dW}}{{dt}}
=(\frac{{dW}}{{dt}})_{T=0}+(\frac{{dW}}{{dt}})_{T\neq0}
\end{equation}
where
\begin{equation}\label{a53}
(\frac{{dW}}{{dt}})_{T=0} = \frac{ e^2v}{2\pi^2 \epsilon_0}\int_{0
}^{ + \infty }d\omega \, \omega \int_{0 }^{ + \infty } {dk}k \,
{\frac{ {{Im(k^2 c^2 \kappa (\omega)-\omega^2 \varepsilon (\omega))}
}}{|k^2 c^2 \kappa (\omega)-\omega^2 \varepsilon
(\omega)|^2}}(1-\frac{\omega^2}{k^2v^2})
\end{equation}
and
\begin{equation}\label{a54}
(\frac{{dW}}{{dt}})_{T\neq0}= \frac{ e^2v}{2\pi^2 \epsilon_0}\int_{0
}^{ + \infty }d\omega
\frac{2\omega}{e^{(\hbar\omega/2k_BT)}-1}\int_{0 }^{ + \infty }
{dk}k \, {\frac{ {{Im(k^2 c^2 \kappa (\omega)-\omega^2 \varepsilon
(\omega))} }}{|k^2 c^2 \kappa (\omega)-\omega^2 \varepsilon
(\omega)|^2}}(1-\frac{\omega^2}{k^2v^2}).
\end{equation}
The last formula differs from the zero-temperature formula only by
the multiplicative factor $2[e^{(\hbar\omega/2k_BT)}-1]^{-1}$ which
has the asymptotic behavior $0$ and $4k_BT/\hbar\omega$ at low
temperatures $k_BT<<\hbar\omega$ and at high temperatures
$k_BT>>\hbar\omega$, respectively. In fact, the Matsubara frequency
sum naturally separates into a term which is temperature independent
and a term containing the Bose-Einstein distribution. In some sense
the replacement of Matsubara frequency sum by an integral, as in
(\ref{a51}), is equivalent to switching from imaginary time
(discrete frequencies in Euclidean space) to real time (continuous
energies in Minkowski space).

A transparent magnetodielectric medium can be considered in principle as a limiting case of a lossy dispersive
medium. In this case using Eq. (\ref{a47}), we obtain
\begin{equation}\label{a55}
\frac{{dW}}{{dt}}=\frac{ e^2v}{4\pi\epsilon_0 c^2 }\sum_j\int_{0 }^{
+ \infty }d\Omega_j\,\Omega_j\,\mu(\Omega_j)\,
\coth(\frac{\hbar\Omega_j}{2k_BT})(1-\frac{c^2}{v^2n^2(\Omega_j)})
\end{equation}
which is the finite temperature generalization of Eq. (\ref{a48}).
It is easy to show that for a nondispersive medium, Eq. (\ref{a48})
tends to the result of \cite{30}.
\section{Quantum Theory}
The classical theory of cherenkov radiation effects is sufficiently accurate in the optical part of the spectrum
\cite{10}. For methodological and physical reasons, it is equally important to consider the quantum theory of
this effects. Quantum theory enables us to derive the classical equation with the appropriate corrections. We
extract the Maxwell equations and constitute relations and the vector potential field operator in the first part
of this section, and in the following we calculate the radiation intensity within a nonrelativistic and
relativistic theory.
\subsection{Canonical quantization}
In the description of the canonical quantization of the
electromagnetic field, we choose the Coulomb gauge ${\rm k} \cdot
\underline{{\rm A}}({\rm k},t) = 0$. In this gauge the vector
potential $\underline{\rm A}$ is a purely transverse field and can
be decomposed along the unit polarization vectors $e_\lambda  ({\rm
k})\,\,\,\,\lambda=1,2 $
\begin{eqnarray} \label{b1}
\underline {\rm A} ({\rm k},t) = \sum\limits_{\lambda  = 1}^2 {\underline A _\lambda  ({\rm k},t)} \,{\rm
e}_\lambda  ({\rm k}).
\end{eqnarray}
The dynamical fields $\underline {\rm X} _\omega$ and $\underline {\rm Y} _\omega$ has both transverse and
longitudinal parts and can be expanded as
\begin{equation} \label{b2}
\underline {\rm X} _\omega  ({\rm k},t) = \sum\limits_{\lambda  = 1}^3 {\underline X _{\omega \lambda } ({\rm
k},t)} \,{\rm e}_\lambda ({\rm k}).
\end{equation}
Furthermore, by using the Lagrange's equation for the scalar potential $\phi$, we find a Lagrangian depending on
a reduced number of dynamical variables in the reciprocal space
\begin{eqnarray} \label{b3}
\underline{L} &=& \frac{1}{2}\sum_\alpha{m_\alpha\rm \dot r}^2_\alpha(t)\nonumber\\
&+& \sum_{\lambda=1}^3 \int_0^\infty {d\omega \int^{'}{d^3{\rm k}(|{\underline {{\dot X}}_{\omega\lambda}} |^2 -
\omega ^2 |{\underline { X}_{\omega\lambda}}|^2 } }+ |{\underline {{ \dot Y}}_{\omega\lambda}}|^2 - \omega ^2
|{\underline {Y}_{\omega\lambda}}|^2)\nonumber\\
&+&\sum_{\lambda=1}^2 \int^{'} {d^3{\rm k}(\varepsilon _0
|{\underline {{ \dot A}}_\lambda}|^2 - \frac{{|{{\rm k} \underline
{A}_\lambda}|^2 }}{{\mu _0 }})} +\int {d^3 {\rm k}}
(\underline{A}_\lambda \underline{J}_\lambda^{*\bot} ({\rm
k},t)+h.c.)\nonumber\\
&+& \sum\limits_{\lambda  = 1}^2 {} \int ^{'}{d^3 {\rm k}} [(
-\underline{ \dot A}_\lambda  \underline{P}_\lambda^{*\bot}({\rm
k},t) +
(\iota{\rm k}\times ( \underline{A}_\lambda  {\rm e}_\lambda ({\rm k})) \cdot \underline{{\rm M}}^* ({\rm k},t)+ h.c.]\nonumber\\
&+& \int {d^3 {\rm k}} (\frac{{(\iota {\rm k} \cdot {\underline{\rm
P}})\underline{\rho} ^* }}{{\varepsilon _0 |k|^2 }} + h.c.) - \int
{d^3 {\rm k}} \frac{{ |\underline{\rho}| ^{2} }}{{\varepsilon _0
|k|^2 }}- \int {d^3 {\rm k}} \frac{{|\iota {\rm k} \cdot
\underline{{\rm P}}|^2}}{{\varepsilon _0 |k|^2 }}.
\end{eqnarray}
The Lagrangian (\ref{b4}) can now be used to obtain the
corresponding canonical conjugate variables of the fields within the
half $\rm k$-space as
\begin{eqnarray} \label{b4}
&& -\underline {D}^\bot  _\lambda  ({\rm k},t) = \frac{{\delta \underline L }}{{\delta (\underline {\dot A}
_\lambda ^* )}} =
\varepsilon _0 \underline {\dot A} _\lambda  ({\rm k},t)-  P_\lambda^{\bot} ({\rm k},t) \\
&& \underline Q _{\omega \lambda } ({\rm k},t) = \frac{{\delta \underline L }}{{\delta (\underline {\dot X}
_{\omega \lambda }^* )}} = \underline {\dot X} _{\omega \lambda } ,\hspace{1cm} \underline \Pi _{\omega \lambda
} ({\rm k},t) = \frac{{\delta \underline L }}{{\delta (\underline {\dot Y} _{\omega \lambda }^*
)}} = \underline {\dot Y} _{\omega \lambda }  \\
&& {\bf p}_\alpha(t) = \frac{{\delta \underline{L}}}{{\delta (\dot
{\bf r}_\alpha )}} = m_\alpha\dot {\bf r}_\alpha + q_\alpha {\bf A}
({\rm r}_\alpha,t).
\end{eqnarray}
Now the fields can be quantized canonically in a standard fashion by demanding equal-time commutation relations
among the variables and their conjugates. For electromagnetic field components, and the dynamical variables of
the external charges, we find respectively
\begin{eqnarray} \label{b5}
&& \left[ {\underline {A} _\lambda ^* ({\rm k},t),-\underline {D}^\bot _{\lambda'} ({\rm k'},t)} \right] = \iota
\hbar \delta _{\lambda \lambda '} \delta ({\rm k} - {\rm k'})
\end{eqnarray}
\begin{eqnarray} \label{b6}
 &&\left[ { r_{\alpha i} (t),p_{\alpha j} (t)} \right] = \iota \hbar \delta _{ij}
\end{eqnarray}
and for the reservoir fields
\begin{equation}\label{b7}
\left[ {\underline {X} _{\omega \lambda }^* ({\rm k},t),\underline Q _{\omega '\lambda '} ({\rm k'},t)} \right]
= \iota \hbar \delta _{\lambda \lambda '} \delta (\omega  - \omega ')\delta ({\rm k} - {\rm k'})
\end{equation}
\begin{equation}\label{b8}
\left[ {\underline {Y} _{\omega \lambda }^* ({\rm k},t),\underline \Pi _{\omega '\lambda '} ({\rm k'},t)}
\right] = \iota \hbar \delta _{\lambda \lambda '} \delta (\omega  - \omega ')\delta ({\rm k} - {\rm k'})
\end{equation}
with all other equal-time commutators being zero. Using the Lagrangian (\ref{b4}) and the expression for the
canonical conjugate variables in (\ref{b5}), we obtain the Hamiltonian of the total system
\begin{eqnarray} \label{b9}
H &=& \sum\limits_{\lambda  = 1}^2\int^{'} d^3 k(\frac{
|\underline{D}_\lambda^\bot - \underline{P}_\lambda^\bot
|^2}{2}+\frac{|{\bf
k}\underline{A}_\lambda|^2}{\mu_0})+\sum_\alpha\frac{({\bf
p}_\alpha-q_\alpha {\bf
A}({\rm r}_\alpha,t))^2}{2m_\alpha} \nonumber \\
&+& \sum_{\lambda=1}^3 \int_0^\infty {d\omega \int^{'} {d^3{\rm
k}(|{\underline {{\dot X}} _{\omega\lambda} }|^2 +\omega ^2
|{\underline { X} _{\omega\lambda} }|^2 } }+ t|{\underline {{ \dot
Y}}_{\omega\lambda} }|^2 + \omega ^2 |{\underline { Y}
_{\omega\lambda}}|^2
)\nonumber\\
&-& \sum\limits_{\lambda  = 1}^2 {} \int ^{'}{d^3 {\rm k}[(\iota{\rm
k}}\times ( \underline{A}_\lambda  {\rm e}_\lambda ({\rm k})) \cdot
\underline{{\rm M}}^* ({\rm k},t) + h.c.]+ \int {d^3 {\rm k}}
\frac{{|\iota {\rm k} \cdot \underline{{\rm
P}}|^2}}{{\varepsilon _0 |k|^2 }}\nonumber\\
&+& \int {d^3 {\rm k}} (\frac{{(-\iota {\rm k} \cdot
\underline{{\rm P}})\underline{\rho} ^* }}{{\varepsilon _0 |k |^2
}} + h.c.) + \int {d^3 {\rm k}} \frac{{|\underline{\rho} | ^2
}}{{\varepsilon _0 |k|^2 }}.
\end{eqnarray}
If we apply Heisenberg equation to the operators $\underline D _\lambda$ and $\underline A _\lambda$, and use
the commutation relation (\ref{b5}), Maxwell equations in the reciprocal space can be obtained as
\begin{eqnarray} \label{b10}
\underline {\dot A} _\lambda  ({\rm k},t) = \frac{\iota }{\hbar }[ {H,\underline A _\lambda  ({\rm k},t)} ] =  -
\frac{{\underline D _\lambda ^ \bot  ({\rm k},t)-\underline P_\lambda ^ \bot  ({\rm k},t)}}{{\varepsilon _0 }}
\end{eqnarray}
\begin{eqnarray} \label{b11}
\underline {\dot D} _\lambda ^ \bot  ({\rm k},t) &=& \frac{\iota }{\hbar }[ {H,\underline D _\lambda ^ \bot
({\rm k},t)}]=\frac{{|k|^2 }}{{\mu _0 }}\underline A _\lambda ({\rm k},t) -{\bf e}_\lambda ({\rm
k})\cdot(\imath{\bf k}\times \underline{{\bf M}}({\rm k},t))- \underline J _\lambda ^ \bot ({\rm k},t)
\end{eqnarray}
Multiplying both sides of these equations by the polarization unit vectors and summing over the polarization
indices we find
\begin{eqnarray} \label{b12}
&& \underline {\rm D} ^ \bot  ({\rm k},t) = \varepsilon _0 \underline {\rm E} ^ \bot  ({\rm k},t)+\underline
{\rm P} ^ \bot ({\rm k},t)
\end{eqnarray}
\begin{eqnarray} \label{b13}
&& \underline {{\rm \dot D}} ^ \bot  ({\rm k},t) = \iota {\rm k} \times \underline {\rm H} ({\rm k},t) -
\underline {\rm J}^\bot ({\rm k},t)
\end{eqnarray}
where $\underline {\rm D}^\bot$ is the transverse displacement field, $\underline {\rm E} ^ \bot   =  -
\underline {{\rm \dot A}}$ is the transverse electric field and $\mu _0 \underline {\rm H} ({\rm k},t) = \iota
{\rm k} \times \underline {\rm A} ({\rm k},t)$ is the magnetic induction field and
\begin{equation}\label{b14}
\underline {\rm J}^\bot ({\rm k},t)=\frac{1}{{2(2\pi )^{{3 \mathord{\left/ {\vphantom {3 2}} \right.
\kern-\nulldelimiterspace} 2}} }}\sum_\alpha q_\alpha({\rm \dot r}_\alpha  e^{ - \iota {\rm k} \cdot {\rm
r}_\alpha}+ e^{ - \iota {\rm k} \cdot {\rm r}_\alpha}{\rm \dot r}_\alpha).
\end{equation}
In the presence of external charges, the longitudinal components of the electric and displacement fields can be
written respectively as
\begin{eqnarray} \label{b15}
&& \underline {\rm E} ^ \|  ({\rm k},t) = -\frac{\hat{k}(\hat{k}\cdot {\bf \underline{P}})}{\epsilon_0} -
\frac{{\iota {\bf k}\underline \rho  ({\rm k},t)}}{{\varepsilon _0 |k|^2 }} \\
&& \underline {\rm D} ^ \|  ({\rm k},t) = \varepsilon _0 \underline {\rm E} ^ \|  ({\rm k},t)+ \underline {\rm
P} ^ \|  ({\rm k},t)=  - \frac{{\iota {\bf k}\underline \rho  ({\rm k},t)}}{{|k|^2 }}.
\end{eqnarray}
If we differentiate the equation (\ref{b12}) with respect to the time variable $t$ and use the equation
(\ref{b13}), we find the quantum counterpart of the equation (\ref{a11}) for the vector potential in the
reciprocal space as
\begin{equation}\label{b16}
\mu_0\epsilon_0\underline {{\rm \ddot A}} ({\rm k},t)+|k|^{ 2} \underline {\rm A} ({\rm k},t)-\mu_0\underline
{{\rm \dot{P}}} ^\bot({\rm k},t)-\mu_0 \imath{\bf k}\times\underline {{\rm M}} ({\rm k},t)= \mu _0 \underline
{\rm J} ^ \bot ({\rm k},t)
\end{equation}
Equation (\ref{b16}) is the Langevin equation for the vector
potential $\bf\underline{A}({\rm k},t)$, wherein, the explicit form
of the electric and magnetic polarization densities of the medium is
known. The quantum Langevin equation can be considered as the basis
of the macroscopic description of a quantum particle coupled to an
environment or a heat bath \cite{31}. Similarly, it is easy to show
that the Heisenberg equation of motion for the external charged
particles is
\begin{equation}\label{b17}
m_\alpha  {\rm \ddot r}_\alpha   = \frac{\iota }{\hbar }[H,{\rm p}_\alpha   - q_\alpha  {\rm A}({\rm r}_\alpha
,t)] = q_\alpha {\rm E}({\rm r}_\alpha  ,t) + \frac{1}{2}q_\alpha  ({\rm \dot r}_\alpha \times {\rm B}({\rm
r}_\alpha  ,t) - {\rm B}({\rm r}_\alpha  ,t) \times {\rm \dot r}_\alpha).
\end{equation}
Using the commutation relations (\ref{b7}), (\ref{b8}) and applying the total Hamiltonian (\ref{b9}), it can be
shown that the combination of the Heisenberg equations of the canonical variables $ {\rm X}(\omega,t) $ and $
{\rm Y}(\omega,t)$ lead to the same equations (\ref{a15}) and (\ref{a16}) with the solutions (\ref{a17}) and
(\ref{a22}), respectively. Now by Substituting the equations (\ref{a17}) and (\ref{a22}) in the integrands of
the equations (\ref{a13}), we find the the electric and magnetic polarization densities of the medium in
reciprocal space
\begin{equation}\label{b18}
\underline{{\rm P}}({\rm k},t) = \epsilon_0\int_0^\infty  {dt\,} \chi_e(t - t'){\bf \underline{E}}({\rm
k},t')+\underline{{\rm P}}^N({\rm k},t)
\end{equation}
\begin{equation}\label{b19}
\underline{{\rm M}}({\rm k},t) = \frac{1}{\mu_0}\int_0^\infty {dt\,} \chi_m(t - t'){\bf \underline{B}}({\rm
k},t')+\underline{{\rm M}}^N({\rm k},t)
\end{equation}
where $ \chi_e$ and $ \chi_m$ are the same electric and magnetic susceptibilities of the medium defined in the
equations (\ref{a19}) and (\ref{a23}), the noises
\begin{eqnarray} \label{b20}
&&\underline{{\rm P}}^N({\rm k},t)=  \int_0^\infty  {d\omega } f(\omega )( { {  \underline {\dot {\bf X}}
_\omega ({\rm k},0)\frac{\sin \omega t}{\omega}}+\underline{{\bf X}}_{\omega}} ({\rm k},0)\cos \omega t ),
\end{eqnarray}
\begin{eqnarray} \label{b21}
&&\underline{{\rm M}}^N({\rm k},t)=  \int_0^\infty  {d\omega } g(\omega )( { {  \underline {\dot {\bf Y}}
_\omega ({\rm k},0)\frac{\sin \omega t}{\omega}}+\underline{{\bf Y}}_{\omega}} ({\rm k},0)\cos \omega t )
\end{eqnarray}
are the electric and magnetic polarization noise densities associated with absorption, with the causal behavior
of the medium, respectively. To facilitate the calculations, let us introduce the following annihilation
operators
\begin{eqnarray} \label{b22}
&& a_\lambda  ({\rm k},t) = \sqrt {\frac{1}{{2\hbar \varepsilon _0 c|k| }}} \left( {\varepsilon _0 c|k|
\underline A _\lambda ({\rm k},t) - \iota \underline D _\lambda ^ \bot  ({\rm k},t)} \right),
\end{eqnarray}
\begin{eqnarray} \label{b23}
&& d_\lambda  ({\rm k},\omega ,t) = \sqrt {\frac{1}{{2\hbar \omega }}} \left( {\omega \underline X _{\omega
\lambda } ({\rm k},t) + \iota \underline Q _{\omega \lambda } ({\rm k},t)} \right),
\end{eqnarray}
\begin{eqnarray} \label{b24}
&& b_\lambda  ({\rm k},\omega ,t) = \sqrt {\frac{1}{{2\hbar \omega }}} \left( {\omega \underline Y _{\omega
\lambda } ({\rm k},t) + \iota \underline \Pi _{\omega \lambda } ({\rm k},t)} \right).
\end{eqnarray}
From equal-time commutation relations for the fields (\ref{b5})-(\ref{b8}), we obtain the following equal-time
commutation relations for the creation and annihilation operators
\begin{eqnarray} \label{b25}
&& \left[ {a_\lambda  ({\rm k},t),a_{\lambda '}^ \dag  ({\rm k'},t)} \right] = \delta _{\lambda \lambda '}
\delta ({\rm k} - {\rm k'})
\end{eqnarray}
\begin{eqnarray} \label{b26}
&& \left[ {d_\lambda  ({\rm k},\omega ,t),d_{\lambda '}^ \dag ({\rm k'},\omega ',t)} \right] = \delta _{\lambda
\lambda '} \delta (\omega  - \omega ')\delta ({\rm k} - {\rm k'})
\end{eqnarray}
\begin{eqnarray} \label{b27}
&& \left[ {b_\lambda  ({\rm k},\omega ,t),b_{\lambda '}^ \dag ({\rm k'},\omega ',t)} \right] = \delta _{\lambda
\lambda '} \delta (\omega  - \omega ')\delta ({\rm k} - {\rm k'})
\end{eqnarray}
The commutation relations (\ref{b25})-(\ref{b27}) in contrast to the
previous relations (\ref{b5})-(\ref{b8}), which were correct only in
the half $\rm k$-space, are now valid in the whole reciprocal space.
Inverting the equations (\ref{b22})-(\ref{b23}), we can write the
canonical variables $\underline {\rm A}$, $\underline {\bf X}
_{\omega }$ and $\underline {\bf Y} _{\omega }$ in terms of the
creation and annihilation operators as
\begin{eqnarray} \label{b28/1}
&& \underline {\rm A} ({\rm k},t) ={\sqrt {\frac{\hbar }{{2\varepsilon _0 c|k| }}} } \sum\limits_{\lambda  =
1}^2
 \left( {a_\lambda
({\rm k},t) + a_\lambda ^ \dag  ( - {\rm k},t)} \right)e_\lambda ({\rm k}),
\end{eqnarray}
\begin{eqnarray} \label{b28/2}
&& \underline {\rm X} _\omega  ({\rm k},t) = \sqrt {\frac{\hbar }{{2\omega }}} \sum\limits_{\lambda  = 1}^3
{\left( {d_\lambda ({\rm k},\omega ,t) + d_\lambda ^ \dag  ( - {\rm k},\omega ,t)} \right)e_\lambda  ({\rm k})},
\end{eqnarray}
\begin{eqnarray} \label{b28/3}
&& \underline {\rm Y} _\omega  ({\rm k},t) = \sqrt {\frac{\hbar }{{2\omega }}} \sum\limits_{\lambda  = 1}^3
{\left( {b_\lambda ({\rm k},\omega ,t) + b_\lambda ^ \dag  ( - {\rm k},\omega ,t)} \right)e_\lambda  ({\rm k})}
\end{eqnarray}
Now by employing the Fourier transforms of these recent relations, the Hamiltonian of the total system
(\ref{b9}), in the real space, can be recast into the final form
\begin{eqnarray} \label{b29}
H &=& \int {d^3 {\rm r}} [ - \frac{{{\rm D}^ \bot  ({\rm r},t) \cdot {\rm P}({\rm r},t)}}{{\varepsilon _0 }} +
\frac{{{\rm P}^2 ({\rm r},t)}}{{2\varepsilon _0 }} - \nabla  \times {\rm A}({\rm r},t)
\cdot {\rm M}({\rm r},t)]\nonumber\\
&+& \sum_\alpha\frac{{\left( {{\rm p}_\alpha - q_\alpha{\rm
A}({\rm r}_\alpha,t)} \right)^2 }}{2m_\alpha}  + \frac{1}{{8\pi
\varepsilon _0 }}\int {d^3 {\rm r}} \int {d^3 {\rm r'}}
\frac{{(\nabla  \cdot {\rm P}({\rm r},t))(\nabla ' \cdot {\rm
P}({\rm r'},t))}}{{|
{{\rm r} - {\rm r'}} |}}\nonumber \\
&-& \frac{1}{{4\pi \varepsilon _0 }}\sum\limits_\alpha  {q_\alpha
} \int {d^3 {\rm r}} \frac{{(\nabla  \cdot {\rm P}({\rm
r},t))}}{{| {{\rm r} - {\rm r}_\alpha  } |}}  + \frac{1}{{8\pi
\varepsilon _0 }}\sum\limits_{\alpha  \ne \beta } {}
\frac{{q_\alpha  q_\beta  }}{{| {{\rm r} - {\rm r}_\alpha }
|}} + H_F + H_e+H_m  \nonumber\\
\end{eqnarray}
where
\begin{eqnarray}\label{b30}
{\rm P}({\rm r},t)=\sum\limits_{\lambda  = 1}^3 {\int_0^\infty {d\omega } } \int {d^3 {\rm k}} \,\sqrt
{\frac{\hbar }{{2\omega }}}
 f(\omega )\left( {d_\lambda ({\rm k},\omega ,t)e^{\imath{\bf k}\cdot{\bf r}}
+h.c.} \right){\bf e}_\lambda ({\rm k}),
\end{eqnarray}
\begin{eqnarray}\label{b31}
{\rm M}({\rm r},t)=\sum\limits_{\lambda  = 1}^3 {\int_0^\infty {d\omega } } \int {d^3 {\rm k}} \,\sqrt
{\frac{\hbar }{{2\omega }}}
 g(\omega )\left( {b_\lambda ({\rm k},\omega ,t)e^{\imath{\bf k}\cdot{\bf r}}
+h.c.} \right){\bf e}_\lambda ({\rm k})
\end{eqnarray}
and
\begin{eqnarray} \label{b32}
&&H_F  = \sum\limits_{\lambda  = 1}^2 {\int {d^3 {\rm k}} \,\,\hbar c|k| \,a_\lambda ^ \dag  ({\rm
k},t)a_\lambda
({\rm k},t)} \\
&& H_e  = \sum\limits_{\lambda  = 1}^3 {\int {d\omega } } \int {d^3 {\rm k}} \,\,\hbar \omega \,\,d_\lambda ^
\dag  ({\rm k},\omega
,t)d_\lambda  ({\rm k},\omega ,t)\\
&& H_m  = \sum\limits_{\lambda = 1}^3 {\int {d\omega } } \int {d^3 {\rm k}} \,\,\hbar \omega \,\,b_\lambda ^
\dag  ({\rm k},\omega ,t)b_\lambda  ({\rm k},\omega ,t)
\end{eqnarray}
are the Hamiltonian of the electromagnetic field and the medium in the normal ordering form.

We now proceed to solve the wave equation (\ref{b16}) along the lines of the classical wave equation (\ref{a28})
using the Laplace transform technique. After some lengthy and elaborated calculations (see appendix), the vector
potential in the large-time limit, i.e. when the medium and electromagnetic field tend to an equilibrium state,
can be obtained as
\begin{eqnarray} \label{b33}
{\rm A}({\rm r},t) &=& \frac{{ - \iota }}{{\varepsilon _0 }}\sum\limits_{\lambda = 1}^2 {} \int {d^3 {\rm k}}
\int {d\omega } \omega \sqrt {\frac{\hbar }{{2(2\pi )^3 \omega }}} f(\omega )\left( {\frac{{d_\lambda ({\rm
k},\omega ,0)e^{ - \iota \omega t} e^{\iota {\rm k} \cdot {\rm r}} }}{{ - \omega ^2 \varepsilon (\omega ) + c^2
k^2 \kappa (\omega )}} - h.c.} \right){\rm e}_\lambda  ({\rm k})\nonumber\\
&+& \frac{\iota }{{\varepsilon _0 }}\sum\limits_{\lambda  = 1}^2 {} \int {d^3 {\rm k}} \int {d\omega }  \sqrt
{\frac{\hbar{|k|^2} }{{2(2\pi )^3 \omega }}}  g(\omega )\left( {\frac{{b_\lambda  ({\rm k},\omega ,0)e^{ - \iota
\omega t} e^{\iota {\rm k} \cdot {\rm r}} }}{{ - \omega ^2 \varepsilon (\omega ) + c^2 k^2 \kappa (\omega )}}
- h.c.} \right){\rm s}_\lambda  ({\rm k})\nonumber\\
\end{eqnarray}
where ${\rm s}_\lambda ({\rm k})=\hat{k}\times {\rm e}_\lambda ({\rm k})$. In the large-time limit, the vector
potential operator will be a function of the medium operators only, i.e, the radiation is due to the medium,
which still satisfies Maxwell's equations, as expected. Also the canonical commutation relation (\ref{b5}) is
preserved in this limit if in addition to the velocity sum rules (\ref{a39}) and (\ref{a40}), which still
legitimate in the quantum domain, the following velocity sum rule for a magnetodielectric medium is also
satisfied (see appendix)
\begin{eqnarray} \label{b34}
\left[ {A^* (k,t), - D^ \bot  (k',t)} \right] &=&\int_0^{ + \infty
} {d\omega } \frac{{\omega ^3 {\mathop{ Im}\nolimits} \varepsilon
(\omega ) - k^2 c^2 \omega {\mathop{ Im}\nolimits} \kappa (\omega
)}}{{|{ - \omega ^2 \varepsilon (\omega ) + c^2 k^2 \kappa
(\omega )} t|^2 }} =\frac{\pi}{2}\nonumber\\
&\Rightarrow&\sum_jRe[\frac{v_g^jv_p^j}{c^2}]=1.
\end{eqnarray}
The form of vector potential operator given in (\ref{b33}) agrees
with previous works, if one replaces the medium annihilation
operators $d_\lambda({\rm k},\omega ,0)$ and $b_\lambda({\rm
k},\omega ,0)$ in the large-time (\ref{b29}) with the diagonalizing
annihilation operators $K_{e,\lambda}({\rm k},\omega)$ and
$K_{m,\lambda}({\rm k},\omega)$, derived by the damped polarization
formalism \cite{24}, \cite{32} and also if one makes similar
replacements for the creation operators $f_\lambda^e({\rm
k},\omega)$ and $f_\lambda^m({\rm k},\omega)$ derived by the
phenomenological formalism \cite{33}, \cite{34}, where again the
same expressions for the field and medium operators are recovered.
\subsection{Nonrelativistic quantum theory of Cherenkov radiation}
We consider a charge particle with mass $m$ and electric charge $e$
uniformly moving in a linear homogeneous magnetodielectric medium
described by the Hamiltonian (\ref{b29}). In fact, we consider a
total system of two noninteracting parts, that is, the free electron
and a system, which consist of the electromagnetic field and the
magnetodielectric medium in interaction. Therefore the Hamiltonian
operator of the total system (\ref{b29}) i.e electromagnetic field,
the medium and the particle, in the large-time limit, can be
rewritten as
\begin{eqnarray} \label{c1}
&&H=H_0+H_{int}\nonumber \\
&&H_0=H_{ele}+H_F\nonumber\\
&&H_{ele}=\frac{{\bf p}^2}{2m}\nonumber\\
&&H_F=:\sum\limits_{\lambda = 1}^3 {\int {d\omega } } \int {d^3 {\rm
k}} \,\,\hbar \omega \,\,\{d_\lambda ^ \dag ({\rm k},\omega
,t)d_\lambda  ({\rm k},\omega ,t)+b_\lambda ^ \dag  ({\rm k},\omega
,t)b_\lambda ({\rm k},\omega ,t)\}:\nonumber\\
&&H_{int}=-\frac{ e{\bf p}\cdot {\bf A}({\bf x},t)}{m}
\end{eqnarray}
where ${\bf x}$ is the position operator of the particle and we have have ignored the term ${\bf A}^2$ since the
Cherenkov radiation can be considered as a first order process in which the number of photons changes by $\pm1$.
Also the direct Coulomb interaction between the electron and the medium has been omitted from the Hamiltonian
(\ref{b29}) since it can give rise to radiative transitions only in third or higher orders.

The moving charged particle with the momentum $\hbar {\rm q}$ has
the quantum state $|{\rm q}\rangle$
\begin{equation}\label{c2}
{\rm p}\left|{{\rm q}} \right\rangle  = \hbar{\rm q}\left| {{\rm
q}} \right\rangle,
\end{equation}
where $\left| {{\rm q}} \right\rangle $ is the momentum eigenvector of the particle which in a coordinate
representation can be written as
\begin{equation}\label{c3}
\left\langle {\rm x} \right.\left| {\rm q} \right\rangle =\psi_{\bf q} ({\rm x})=\frac{1}{{(2\pi )^{{3
\mathord{\left/
 {\vphantom {3 2}} \right.
 \kern-\nulldelimiterspace} 2}} }}e^{\iota \,{\rm q} \cdot {\rm x}},
\end{equation}
therefore the Hamiltonian $H_{ele}={{{\rm p}^2 } \mathord{\left/
 {\vphantom {{{\rm p}^2 } {2m}}} \right.
 \kern-\nulldelimiterspace} {2m}}
$ for a free particle has the eigenvector $\left| {{\rm q}}
\right\rangle $ with the energy eigenvalues $E_{\rm q}  = {{\hbar ^{
2} {\rm q}^2 } \mathord{\left/
 {\vphantom {{\hbar ^{\rm 2} {\rm q}^2 } {2m}}} \right.
 \kern-\nulldelimiterspace} {2m}}
$. The unperturbed Hamiltonian $H_0=H_{ele}+H_F$ has the eigenstate
\begin{equation}\label{d6/3}
\mid ele+ rad\rangle=\mid ele\rangle \mid  rad\rangle\nonumber
\end{equation}
which are the direct product of the eigenstates of $H_{ele}$ and $H_F$. In order to separate out the emission of
Cherenkov radiation from various other processes which might occur, such as, for example, ionization, emission
of bremsstrahlung, etc., we restrict our attention to first order transitions. Therefore We apply quantum
mechanical perturbation theory up to the first order approximation to treat the transition probability per unit
time for a free particle of momentum $\hbar {\rm q}$ to emit a photon of momentum $\hbar {\rm k}$ and energy
$\hbar\omega$ thereby changing its momentum to $\hbar ({\rm q}-{\rm k})$ as following
\begin{equation}\label{c4}
\Gamma _{{\rm q} \to {\rm q} - {\rm k}}  = \frac{{2\pi }}{\hbar
}\left|\langle 1_{\bf k} \mid \langle {\bf q}-{\bf k} \mid H_{int
}\mid {\bf q}\rangle \mid  0\rangle \right|^2 \delta (\frac{{\hbar
^2 {\rm q}^2 }}{{2m}} - \frac{{\hbar ^2 }}{{2m}}\left| {{\rm q} -
{\rm k}} \right|^2  - \hbar \omega )
\end{equation}
where the states $\mid  0\rangle$ and $\mid 1_{\bf k} \rangle$
present the vacuum state of the electromagnetic field and the
excited state of the electromagnetic field with a single photon with
wave vector ${\rm k}$ and frequency $\omega$, respectively. The
argument of the Dirac $\delta$ function displays the conservation of
energy and the square of the matrix element of the equation
(\ref{c4}) is obtained by using Eqs. (\ref{b33}) and (\ref{c1}) as
\begin{eqnarray}\label{c5}
\left|\langle 1_{\bf k} \mid \langle {\bf q}-{\bf k} \mid H_{int
}\mid {\bf q}\rangle \mid  0\rangle \right|^2 &=& \frac{{\hbar e^2
}}{{16\pi ^3 \varepsilon _0^2 m^2 \omega }}\{ \left| {\beta (\omega
,| k |)} \right|^2 \left| {\left\langle {{\rm q} - {\rm k}}
\right|e^{ - \iota \,{\rm k} \cdot {\rm x}} {\rm p} \cdot {\rm
e}_\lambda ({\rm k})\left| {\rm q} \right\rangle } \right|^2
\nonumber\\
& +& k^2 \left| {\gamma (\omega ,\left| k \right|)} \right|^2 \left| {\left\langle {{\rm q} - {\rm k}}
\right|e^{ - \iota \,{\rm k} \cdot {\rm x}} {\rm p} \cdot {\rm s}_\lambda  ({\rm k})\left| {\rm q} \right\rangle
} \right|^2 \}
\end{eqnarray}
by using Eqs. (\ref{c3}) and (\ref{c2}) the matrix elements in above equation are just $\hbar{\bf k}\cdot{\rm
e}_\lambda  ({\rm k})$ and $\hbar{\bf k}\cdot{\rm s}_\lambda  ({\rm k})$, respectively. Let $\theta$ be the
angle between ${\rm q}$ and ${\rm k}$ and let ${\rm v} = {{\hbar {\rm q}} \mathord{\left/
 {\vphantom {{\hbar {\rm q}} m}} \right. \kern-\nulldelimiterspace} m}$ be the particle velocity, we find
\begin{equation}\label{c6}
\Gamma _{{\rm q} \to {\rm q} - {\rm k}}  = \frac{{e^2 v(1 - \cos ^2 \theta )}}{{4\pi ^3 \varepsilon _0 \hbar
k}}\left( {\frac{{\omega ^2 {\mathop{ Im}\nolimits} \varepsilon (\omega ) - k^2 c^2 {\mathop{ Im}\nolimits}
\kappa (\omega )}}{{\left| { - \omega ^2 \varepsilon (\omega ) + k^2 c^2 \kappa (\omega )} \right|^2 }}}
\right)\delta (\cos \theta  - \frac{\omega }{{kv}}(1 + \frac{{\hbar k^2 }}{{2m\omega }})),
\end{equation}
therefore, photon is emitted at an angle to the path of the particle given by
\begin{equation}\label{c7}
\cos \theta  = \frac{\omega }{{kv}}(1 + \frac{{\hbar k^2 }}{{2m\omega }}).
\end{equation}
If the energy of the photon $\hbar\omega$ is much less than the rest mass of the particle $mc^2$ then this is
approximately Eq.(\ref{a46/1}) which gives the classical Cherenkov angle. The total energy radiated per unit
time is found to be
\begin{eqnarray}\label{c8}
\frac{dW}{dt}&=&\frac{{e^2 v}}{{2\pi ^2 \varepsilon _0 }}\sum\limits_{\lambda = 1} ^2 \int {d^3 {\bf k}}
\int_0^{ + \infty } {d\omega } \hbar \omega \Gamma _{{\rm q} \to {\rm q}-
{\rm k}}\nonumber\\
&=&\frac{{e^2 v}}{{2\pi ^2 \varepsilon _0 }} \int_0^{ + \infty }
kdk\int_0^{ + \infty }\omega d\omega [1 - \frac{\omega^2 }{k^2v^2}(1
+ \frac{{\hbar k^2 }}{{2m\omega }})^2]{\mathop{Im}\nolimits} \left(
{\frac{1}{{ - \omega ^2
\varepsilon (\omega ) + k^2 c^2 \kappa (\omega )}}} \right)\nonumber\\
\end{eqnarray}
We note that the integration on the azimuthal angle is trivial. The integration on polar angle is done with the
help of the Dirac $\delta$ function in Eq. (\ref{c8}) and
\begin{equation}\label{c9}
\left( {\frac{{\omega ^2 {\mathop{ Im}\nolimits} \varepsilon (\omega ) - k^2 c^2 {\mathop{ Im}\nolimits} \kappa
(\omega )}}{{\left| { - \omega ^2 \varepsilon (\omega ) + k^2 c^2 \kappa (\omega )} \right|^2 }}}
\right)={\mathop{Im}\nolimits} \left( {\frac{1}{{ - \omega ^2 \varepsilon (\omega ) + k^2 c^2 \kappa (\omega
)}}} \right).
\end{equation}
It is easily shown that Eq. (\ref{c8}) is consistent with the result
of \cite {35} and in the classical limit as $\hbar\longrightarrow0$,
reduces to the classical results (\ref{a46}).
\subsection{Relativistic quantum theory of Cherenkov radiation}
The description of particles used in the preceding section is valid
only when the particles are moving at velocities small compared to
the velocity of light. The preceding formalism must be generalized
somehow to describe the relativistic moving particles, that is, we
should drive the Dirac equation for external particles embedded in
the magnetodielectric medium. For this purpose, we substitute the
following Lagrangian for external particle instead of the Lagrangian
(\ref{a4}) \cite {36}, \cite {37}
\begin{eqnarray} \label{d1}
L_q & =& \frac{{\iota \hbar c}}{2}\int {d^3 {\rm x}} \left[ {\sum\limits_{\mu = 0}^3 {\left( {\bar \psi ({\rm
x},t)\gamma ^\mu \frac{{\partial \psi ({\rm x},t)}}{{\partial x^\mu  }} - \frac{{\partial \bar \psi ({\rm
x},t)}}{{\partial x^\mu  }}\gamma
^\mu  \psi ({\rm x},t)} \right)} } \right] - mc^2 \bar \psi \psi \nonumber\\
&+&  e\int {d^3 {\rm x}} \left[ {\sum\limits_{j = 1}^3 {(c\bar \psi ({\rm x},t)\gamma ^j \psi ({\rm x},t){\rm
A}^j ({\rm x},t))}  - \bar \psi ({\rm x},t)\gamma ^0 \psi
({\rm x},t)\varphi ({\rm x},t)} \right]\nonumber\\
\end{eqnarray}
where $\gamma^\mu , \mu=0,...,3$ are the Dirac matrices with $\gamma ^0 = \beta $, $\gamma ^j = \beta \alpha _j
$ and $\bar \psi  = \psi ^\dag \beta$. In a standard representation we have
\begin{equation}\label{d2}
\beta  = \left( {\begin{array}{*{20}c}
   I & 0  \\
   0 & { - I}  \\
\end{array}} \right),\,\alpha _j  = \left( {\begin{array}{*{20}c}
   0 & {\sigma _j }  \\
   {\sigma _j } & 0  \\
\end{array}} \right)
\end{equation}
where $\sigma _j, j=1,2,3$ are Pauli spin matrices and $I$ is the unit matrix. It is worth mentioning that the
Lagrangian of the magnetodielectric medium (\ref{a8}) need not to be written in a covariant form since the
medium is at rest and a non relativistic description is enough although it can be written in a covariant form
straightforwardly. We proceed along the lines of the preceding section and instead of the canonical momentum of
the particle ${\bf p}_\alpha$ we define the canonical conjugate variable of the Dirac particle
$\imath\hbar\psi^\dag$ as
\begin{equation}\label{d2/1}
\frac{\imath\hbar\psi^\dag}{2}=\frac{\partial L}{\partial
\dot{\psi}}.
\end{equation}
The quantization procedure for the Dirac field can be achieved by imposing equal-time anticommutation relation
among the field components
\begin{equation}\label{d2/2}
\{\psi_\alpha({\bf x},t),\psi^\dag_\beta({\bf x'},t)\}=\delta_{\alpha\beta}\delta({\bf x}-{\bf x'})
\end{equation}
together with $\{\psi_\alpha({\bf x},t),\psi_\beta({\bf x'},t)\}=0$. Here we have chosen the anticommutation
relation, since we are developing a theory of particles that obey Fermi-Dirac statistics. Using the Lagrangian
(\ref{b3}) and (\ref{d1}) and the expression for the canonical conjugate variables in (\ref{b5})-(\ref{b8}) and
(\ref{d2/2}), we obtain the Hamiltonian of the total system
\begin{eqnarray} \label{d3}
H &=& \int {d^3 {\rm x}\,} ( - \iota \hbar c\psi ^\dag ({\rm x},t){\boldsymbol{ \alpha }} \cdot \nabla \psi
({\rm x},t) + mc^2 \psi ^\dag ({\rm x},t)\beta \psi ({\rm x},t))
\nonumber \\
&+& \sum_{\lambda=1}^3 \int_0^\infty {d\omega \int^{'} {d^3{\rm k}(\left| {\underline {{\dot X}}
_{\omega\lambda} } \right|^2 +\omega ^2 \left| {\underline { X} _{\omega\lambda} } \right|^2 } }+ \left|
{\underline {{ \dot Y}}_{\omega\lambda} } \right|^2 + \omega ^2 \left| {\underline { Y} _{\omega\lambda}  }
\right|^2
)\nonumber\\
&-& \sum\limits_{\lambda  = 1}^2 {} \int ^{'}{d^3 {\rm k}[(\iota{\rm
k}}\times ( \underline{A}_\lambda  {\rm e}_\lambda ({\rm k})) \cdot
\underline{{\rm M}}^* ({\rm k},t) + h.c.]+ \int^{'} {d^3 {\rm k}}
\frac{{|\iota {\rm k} \cdot \underline{{\rm
P}}|^2}}{{\varepsilon _0 \left| k \right|^2 }}\nonumber\\
&+& \int^{'} {d^3 {\rm k}} (\frac{{(-\iota {\rm k} \cdot
\underline{{\rm P}})\underline{\rho} ^* }}{{\varepsilon _0 \left| k
\right|^2
}} + h.c.) + \int^{'} {d^3 {\rm k}} \frac{{|\underline{\rho}| ^2 }}{{\varepsilon _0 \left| k \right|^2 }}\nonumber\\
&-& \int^{'}{d^3 {\rm k}} (\underline{{\rm J}}^{\rm *} ({\rm k},t)
\cdot \underline{{\rm A}}({\rm k},t) + h.c.)+\sum\limits_{\lambda  =
1}^2\int^{'} d^3 k(\frac{ |\underline{D}_\lambda^\bot -
\underline{P}_\lambda^\bot |^2}{2}.
\end{eqnarray}
It is easily shown that the Heisenberg equation for the dynamic variable of the electromagnetic and medium
fields lead to the same constitute equations (\ref{b18})-(\ref{b19}) and also Maxwell equations
(\ref{b12})-(\ref{b15}) in the reciprocal space. The Fourier transforms of the external current and charge
densities are defined by
\begin{eqnarray}\label{d3/1}
&&\underline{{\rm J}}({\rm k},t) = \frac{{ec}}{{(2\pi )^{{3
\mathord{\left/ {\vphantom {3 2}} \right. \kern-\nulldelimiterspace}
2}} }}\int {d^3 {\rm x}\,} \psi ^\dag ({\rm x},t){\boldsymbol
{\alpha} }\,\psi ({\rm x},t)e^{ -
\iota {\rm k} \cdot {\rm x}} \\
&& \underline{\rho} ({\rm k},t) = \frac{e}{{(2\pi )^{{3
\mathord{\left/ {\vphantom {3 2}} \right. \kern-\nulldelimiterspace}
2}} }}\int {d^3 {\rm x}\,} \psi ^\dag ({\rm x},t)\,\psi ({\rm
x},t)e^{ - \iota {\rm k} \cdot {\rm x}}.
\end{eqnarray}
If we apply Heisenberg equation to the Dirac fields $\psi ({\rm x},t)$ and make use of the anticommutation
relation (\ref{d2/2}), Dirac equation in the presence of the electromagnetic field can be obtained as
\begin{equation}\label{d4}
\iota \hbar \dot \psi ({\rm x},t) =   -c{\boldsymbol{ \alpha }} \cdot ( \iota \hbar\nabla +e{\bf A}({\rm
x},t))\psi({\rm x},t) + e\varphi({\rm x},t)+mc^2 \beta \psi ({\rm x},t),
\end{equation}
where
\begin{equation}\label{d4/1}
\varphi({\rm x},t)=\frac{1}{{(2\pi )^{{3 \mathord{\left/ {\vphantom
{3 2}} \right. \kern-\nulldelimiterspace} 2}} }}\int {d^3 {\rm k}\,}
(\frac{{ - \iota {\rm k} \cdot \underline{{\rm P}}({\rm
k},t)}}{{\varepsilon _0 { k}^{ 2} }} + \frac{{\underline{\rho} ({\rm
k},t)}}{{\varepsilon _0 { k}^{ 2} }})e^{\iota {\rm k} \cdot {\rm
x}},
\end{equation}
is the scalar potential of the electromagnetic field defined in Eq. (\ref{a12}). We now expand the Dirac field
$\psi({\rm x},t)$ in eigenfunctions of the Dirac equation in the absence of the electromagnetic field
\begin{equation}\label{d5}
\psi ({\rm x},t) = \frac{1}{{(2\pi )^{{3 \mathord{\left/ {\vphantom {3 2}} \right. \kern-\nulldelimiterspace}
2}} }}\sum\limits_{\mu  = 1}^4 \int {d^3 {\rm q}\,} c_\mu  ({\rm q},t)\psi _\mu  ({\rm q})
\end{equation}
where $\psi _\mu  ({\rm q}) = u_\mu  ({\rm q})e^{\iota {\rm q} \cdot
{\rm x}} $ and $u_\mu  ({\rm q})$ are four-component spinors of the
Dirac equation with eigenvalues $E_{{\rm q}} = \pm \sqrt {\hbar ^2
c^2 {\rm q}^2 + m^2 c^4 }$ with the normalization condition
$u^\dag_\mu({\rm q}) u_\nu({\rm q})=\delta_{\mu\nu}$ \cite{28} and
$c_\mu ({\rm q},t)$ are the annihilation operators of the particle
with momentum $\hbar{\bf q}$. By substituting Eqs. (\ref{d5}) and
(\ref{b33}) in (\ref{d3}), the Hamiltonian of the total system in
the large-time limit can be recast in the following form
\begin{eqnarray} \label{d6}
H&=&H_0+H_{int}\nonumber \\
H_0&=&H_{ele}+H_F\nonumber\\
\nonumber
\end{eqnarray}
\begin{equation}\label{d6/1}
H_{ele} = \sum\limits_{\mu  = 1}^4 {} \int {d^3 {\rm q}\,} E_{{\rm
q} } c_\mu ^\dag ({\rm q},t)c_\mu  ({\rm q},t),
\end{equation}
\begin{eqnarray}\label{d6/2}
H_F &=& :\sum_{\lambda = 1}^3 \int d\omega\int d^3 {\rm
k}\hbar\omega d_{\lambda}^{\dag} ({\rm k},\omega,t)d_\lambda ({\rm
k},\omega,t)+b_\lambda ^\dag ({\rm k},\omega,t)b_\lambda({\rm
k},\omega,t)\}:\nonumber\\
H_{int}&=&- \int {d^3 {\rm k}} (\underline{{\rm J}}^{\rm *} ({\rm
k},t) \cdot \underline{{\rm A}}({\rm k},t) + h.c.) =  - ec\int {d^3
{\rm x}}\int {d^3 {\rm k}}\, \psi ^* {\boldsymbol{\alpha}} \cdot \underline{{\rm A}}({\rm k},t)\psi \nonumber\\
&=&\frac{{\iota ce}}{\varepsilon_0 }\sum\limits_{\lambda  = 1}^2
\sum\limits_{\mu ,\mu ' = 1}^4  \int {d^3 {\rm k}} \int {d^3 {\rm
q}} \int_0^\infty  {d\omega } \sqrt {\frac{{\hbar \omega }}{{2(2\pi
)^3 }}} f(\omega )\nonumber\\
&\times &(\frac{{u_\mu ^\dag ({\rm q})\boldsymbol{\alpha} \cdot {\rm
e}_\lambda ({\rm k})u_{\mu '} ({\rm q} - {\rm k})}}{{ - \omega ^2
\varepsilon (\omega ) + c^2 k^2 \kappa (\omega )}}c_\mu ^\dag ({\rm
q})c_{\mu '} ({\rm q} - {\rm k})d_\lambda  {\rm (k},\omega
,0)e^{-\imath\omega t}-h.c.) \nonumber\\
&-&\frac{{\iota ce}}{\varepsilon_0 }\sum\limits_{\lambda  = 1}^2 {} \sum\limits_{\mu ,\mu ' = 1}^4 {} \int {d^3
{\rm k}} \int {d^3 {\rm q}} \int_0^\infty  {d\omega } \sqrt {\frac{{\hbar \left| k \right|^2
}}{{2(2\pi )^3 \omega }}} g(\omega )\nonumber\\
&\times &(\frac{{u_\mu ^\dag ({\rm q})\boldsymbol{\alpha} \cdot {\rm s}_\lambda ({\rm k})u_{\mu '} ({\rm q} -
{\rm k})}}{{ - \omega ^2 \varepsilon (\omega ) + c^2 k^2 \kappa (\omega )}}c_\mu ^\dag ({\rm q})c_{\mu '} ({\rm
q} - {\rm k})b_\lambda {\rm (k},\omega ,0)e^{-\imath\omega t} - h.c.).\nonumber\\
\end{eqnarray}
The unperturbed Hamiltonian $H_0=H_{ele}+H_F$ has the eigenstate
\begin{equation}\label{d6/3}
\mid ele+ rad\rangle=\mid ele\rangle \mid  rad\rangle\nonumber
\end{equation}
which are the direct product of the eigenstates of $H_{ele}$ and
$H_F$. we again apply first order perturbation theory to treat the
transition probability per unit time for a free Dirac particle of
momentum $\hbar {\rm q}$ to emit a photon of momentum $\hbar {\rm
k}$ and energy $\hbar\omega$ thereby changing its momentum to $\hbar
({\rm q}-{\rm k})$
\begin{equation}\label{d7}
\Gamma _{{\rm q} \to {\rm q} - {\rm k}} = \frac{{2\pi }}{\hbar
}\left| \langle 1_{\bf k} \mid \langle {\bf q}-{\bf k} \mid H_{int
}\mid {\bf q}\rangle \mid  0\rangle \right|^2 \delta (\sqrt {\hbar
^2 c^2 {\rm q}^2 + m^2 c^4 } - \sqrt {\hbar ^2 c^2 |{{\rm q} - {\rm
k}}|^2 + m^2 c^4 } - \hbar\omega )
\end{equation}
where the argument of the Dirac $\delta$ function displays the
conservation of energy. We may proceed to calculate the energy loss
per unit time as we did in the previous section but with a little
modification here. The sum over final states must include a sum over
the final spin states of the particle with positive energy
$\mu=1,2$, also we average over the initial spin states
\begin{equation}\label{d8}
\frac{dW}{dt}= \frac{1}{2}\sum\limits_{\lambda  = 1}^2 \sum\limits_{\mu ,\mu ' = 1}^2\int d^3{\bf
k}\int_0^\infty d\omega \hbar\omega \Gamma _{{\rm q} \to {\rm q} - {\rm k}}.
\end{equation}
In order to calculate the above equation we must evaluate the following sums
\begin{eqnarray}\label{d9}
&&S = \frac{1}{2}\sum\limits_{\lambda  = 1}^2  \sum\limits_{\mu ,\mu ' = 1}^2 \left| {u_\mu ^\dag ({\rm
q})\alpha  \cdot {\rm e}_\lambda
({\rm k})u_{\mu '} ({\rm q} - {\rm k})} \right|^2  \\
&&S'=  \frac{1}{2}\sum\limits_{\lambda  = 1}^2  \sum\limits_{\mu ,\mu ' = 1}^2 \left| {u_\mu ^\dag ({\rm
q})\alpha  \cdot {\rm s}_\lambda  ({\rm k})u_{\mu '} ({\rm q} - {\rm k})} \right|^2.
\end{eqnarray}
For this purpose we introduce the annihilation operators \cite {38}
\begin{eqnarray}\label{d10}
&&\Lambda ({\rm q}) = \frac{{c{\rm \alpha } \cdot {\rm p} + \beta mc^2  + \left| {E_{\rm q} } \right|}}{{2\left|
{E_{\rm q} }
\right|}} \\
&&\Lambda ({\rm q} - {\rm k}) = \frac{{c{\rm \alpha } \cdot ({\rm q} - {\rm k}) + \beta mc^2  + \left| {E_{{\rm
q} - {\rm k}} } \right|}}{{2\left| {E_{{\rm q} - {\rm k}} } \right|}},
\end{eqnarray}
and we obtain
\begin{eqnarray}\label{d11}
S&=&\frac{1}{8}tr[(\alpha  \cdot {\rm e}_\lambda  ({\rm k}))\Lambda ({\rm q} - {\rm k})(\alpha  \cdot {\rm
e}_\lambda  ({\rm k}))\Lambda ({\rm q})]\nonumber\\
&=&\frac{1}{2}\{ 1 - \frac{{m^2 c^4 }}{{\left| {E_{\rm q} }
\right|\left| {E_{{\rm q} - {\rm k}} } \right|}} + \frac{{2({\rm e}_\lambda  ({\rm k}) \cdot {\rm v}_1 )^2
}}{{c^2 }}  - \frac{{{\rm v}_1  \cdot {\rm
v}_2 }}{{c^2 }}\}\\
S'&= &\frac{1}{8}tr[(\alpha  \cdot {\rm s}_\lambda ({\rm k}))\Lambda ({\rm q} - {\rm k})(\alpha  \cdot {\rm
s}_\lambda ({\rm k}))\Lambda
({\rm q})]\nonumber\\
&=& \frac{1}{2}\{ 1 - \frac{{m^2 c^4 }}{{\left| {E_{\rm q} } \right|\left| {E_{{\rm q} - {\rm k}} } \right|}} +
\frac{{2({\rm s}_\lambda  ({\rm k}) \cdot {\rm v}_1 )^2 }}{{c^2 }}  - \frac{{{\rm v}_1  \cdot {\rm v}_2 }}{{c^2
}}\},
\end{eqnarray}
where we have used ${\rm v} = {{\hbar c^2 {\rm q}} \mathord{\left/
{\vphantom {{c^2 {\rm q}} {E_{\rm q} }}} \right.
\kern-\nulldelimiterspace} {E_{\rm q} }}$ and ${\rm v}_1$ and ${\rm
v}_2$ are the velocities before an after the emission of the photon
respectively. The sum over polarizations can be carried out as in
Eq. (\ref{a46}). The result is
\begin{equation}\label{d12}
S = S'  = \frac{{{\rm v}_1^2 }}{{c^2 }}(1 - \cos ^2 \theta ) + \frac{1}{2}\{ 1 - \sqrt {(1 - {{{\rm v}_1^2 }
\mathord{\left/ {\vphantom {{{\rm v}_1^2 } {c^2 }}} \right. \kern-\nulldelimiterspace} {c^2 }})(1 - {{{\rm
v}_2^2 } \mathord{\left/ {\vphantom {{{\rm v}_2^2 } {c^2 }}} \right. \kern-\nulldelimiterspace} {c^2 }})}  -
\frac{{{\rm v}_1  \cdot {\rm v}_2 }}{{c^2 }}\}
\end{equation}
where again $\theta$ is the angle between ${\rm q}$ and ${\rm k}$ given by
\begin{equation}\label{d13}
cos\theta=\frac{\omega }{{vk}}[1 + \frac{{\hbar \omega }}{{2mc^2 }}(\frac{{k^2 c^2 }}{{\omega ^2 }} - 1)\sqrt {1
- \frac{{v^2 }}{{c^2 }}} ].
\end{equation}
We mention here that in the classical theory only the first term
occurs on the right-hand side of the equation (\ref{d13}) and the
second and third terms are a consequence of non relativistic and
relativistic quantum theory respectively. These terms are small
since the wave length of the electron is much smaller than the
photon's wave length \cite {39}-\cite {40}. Also the second term in
Eq. (\ref{d12}) is a small correction to the result we found in the
preceding section. We neglect this term and the rest of the
calculation is similar to what presented in the preceding section.
The only difference here is that Eq. (\ref{d13}) must be used
instead of Eq. (\ref{c7}). The result is
\begin{eqnarray}\label{d14}
\frac{dW}{dt} &=& \frac{{e^2 v}}{{2\pi ^2 \varepsilon _0 }} \int_0^{
+ \infty }  dkk\int_0^{ + \infty } \,d\omega \omega (1 -
\frac{\omega^2 }{{v^2k^2}}[1 + \frac{{\hbar \omega }}{{2mc^2
}}(\frac{{k^2 c^2 }}{{\omega ^2 }} -
1)\sqrt {1 - \frac{{v^2 }}{{c^2 }}} ]^2)\nonumber\\
&\times &{\mathop{Im}\nolimits} \left( {\frac{1}{{ - \omega ^2
\varepsilon (\omega ) + k^2 c^2 \kappa (\omega )}}} \right).
\end{eqnarray}
Comparing this expression and Eq. (\ref{c8}) with its classical counterpart, Eq. (\ref{a46}), it is seen that
the only difference is in the argument of the $\delta$ function. In fact, we note that the integrations over $k$
and $\omega$ in classical equation (\ref{a46}) diverge but its nonrelativistic and relativistic counterpart that
is Eqs. (\ref{c8}) and (\ref{d14}) have no divergent behavior. To appreciate the physical significance of this
subject, we consider a transparent and nondispersive magnetodielectric medium. We note that Eqs. (\ref{c7}) and
(\ref{d13}) for $0<v<c$ and $n>1$ provide a cutoff in frequency $\omega<\omega_c^{{nonrel},{rel}}$ which in non-
and relativistic regimes are respectively given by
\begin{equation}\label{d14/2}
\omega_c^{nonrel}=\frac{2mc^2(n\beta-1)}{\hbar n^2}
\end{equation}
and
\begin{equation}\label{d14/3}
\omega_c^{rel}=\frac{2mc^2(n\beta-1)}{\hbar (n^2-1)\sqrt{1-\beta^2}}
\end{equation}
where $\beta=v/c$. These cutoffs are still bounded above by the electron energy $mc^2$ or
$mc^2/\sqrt{1-\beta^2}$, respectively. Thus, we can use one of these cutoffs for the range of integrations to
obtain a physically acceptable result for the classical radiation intensity.

It is easy to show that in a transparent magnetodielectric medium Eq. (\ref{d14}) becomes
\begin{eqnarray}\label{d15}
\frac{dW}{dt}=\frac{ e^2v}{4\pi\epsilon_0 c^2 }\sum_j\int_{0 }^{ + \infty }d\Omega_j \Omega_j \mu(\Omega_j) (1 -
\frac{c^2 }{{n^2v^2}}[1 + \frac{{\hbar \Omega_j }}{{2mc^2 }}(n^2 - 1)\sqrt {1 - \frac{{v^2 }}{{c^2 }}} ]^2),\nonumber\\
\end{eqnarray}
which tends to the correct relation in the classical and extreme relativistic limits.
\subsection{Finite temperature Cherenkov radiation in quantum regime}
Our considerations so far have been applied to zero temperature. The
inclusion of temperature may be done in the usual manner \cite{27}.
In this case the transition probability (\ref{d7}) for a free Dirac
particle of momentum $\hbar {\rm q}$ to emit a photon of momentum
$\hbar {\rm k}$ and energy $\hbar\omega$ thereby changing its
momentum to $\hbar ({\rm q}-{\rm k})$ is obtained as
\begin{equation}\label{d16}
\Gamma _{{\rm q} \to {\rm q} - {\rm k}} = \frac{{2\pi }}{\hbar
}\left| H_{int } \right|^2 (N_{\bf k}+1)(1-n_F({\rm q} - {\rm
k}))\delta (\sqrt {\hbar ^2 c^2 {\rm q}^2 + m^2 c^4 } - \sqrt {\hbar
^2 c^2 |{{\rm q} - {\rm k}}|^2 + m^2 c^4 } - \hbar\omega ),
\end{equation}
where
\begin{eqnarray}\label{d17}
N_{\bf k}=\frac{1}{e^{\hbar\omega/k_BT}-1}
\end{eqnarray}
and
\begin{equation}\label{a17/1}
n_F({\rm q} - {\rm k})=\frac{1}{e^{{\sqrt {\hbar ^2 c^2 |{{\rm q} -
{\rm k}}|^2 + m^2 c^4 }/k_BT}}+1}.
\end{equation}
Here the factor $N_{\bf k}+1 $ comes from the phonon creation
operator for transition from the initial state with $n_{\bf k}$
photons to the final state with $n_{\bf k}+1$ photons. In fact, we
use the thermal average of $n_{\bf k}$ which is $N_{\bf k} $. This
photon emission process take place as stimulated and spontaneous
emission. Similarly, the factor $1-n_F({\rm q} - {\rm k})$ is the
probability that the electron state ${\rm q} - {\rm k}$ is empty, so
that the operator $c_\mu^\dag({\rm q} - {\rm k})$ can create an
electron in that state. There is also a factor $n_F({\rm q})$, which
is the probability that ${\rm q}$ is occupied with an electron that
is unity. But in calculating the total energy radiated as Cherenkov
radiation, we have to omit some important processes. They arise from
other electrons in the system with the same spin state. These
electrons can not be found in the state ${\rm q}$ since our electron
is occupying it already. Thus the other electrons of the system will
reduce the transition probability (\ref{d16})and we must subtract
the transition probability due to the presence of other electrons
from Eq. (\ref{d16}) \cite{27}. Therefore by using Eqs.
(\ref{d8})-(\ref{d14}), the total energy radiated in finite
temperature is found to be
\begin{eqnarray}\label{d18}
\frac{dW}{dt} &=& \frac{{e^2 v}}{{2\pi ^2 \varepsilon _0 }} \int_0^{
+ \infty }  dkk\int_0^{ + \infty } \,d\omega \omega (1 -
\frac{\omega^2 }{{v^2k^2}}[1 + \frac{{\hbar \omega }}{{2mc^2
}}(\frac{{k^2 c^2 }}{{\omega ^2 }} -
1)\sqrt {1 - \frac{{v^2 }}{{c^2 }}} ]^2)\nonumber\\
&\times &{\mathop{Im}\nolimits} \left( {\frac{1}{{ - \omega ^2
\varepsilon (\omega ) + k^2 c^2 \kappa (\omega )}}}
\right)F_T(\omega)
\end{eqnarray}
where
\begin{eqnarray}\label{d19}
F_T(\omega)&=&(N_{\bf k}+1)(1-n_F({\rm q} - {\rm k}))-N_{\bf k}n_F({\rm q} - {\rm k})\nonumber \\
&=&\frac{e^{\hbar\omega/k_BT}}{e^{\hbar\omega/k_BT}-1}[\frac{e^{|E_{\bf
q}-\hbar\omega|/k_BT}-e^{-\hbar\omega/k_BT}}{e^{|E_{\bf
q}-\hbar\omega|/k_BT}+1}].
\end{eqnarray}
The last equation differs from the zero temperature Equation (\ref{d14}) only by the multiplicative factor
$F_T(\omega)$ which has the following asymptotic behavior at low temperatures $k_BT<<\hbar\omega$ and at high
temperatures $k_BT>>E_{\bf q}$ respectively:
\begin{eqnarray}\label{d20}
F_T(\omega)&\sim& 1 \nonumber\\
F_T(\omega)&\sim& \frac{E_{\bf q}}{2\hbar\omega}= \frac{mc^2}{2\hbar\omega\sqrt{1-{(v/ c)}^2}}.\nonumber
\end{eqnarray}
It is easy to show that in a transparent magnetodielectric medium, the finite temperature generalization of the
equation (\ref{d14}) becomes
\begin{eqnarray}\label{d21}
\frac{dW}{dt}=\frac{ e^2v}{4\pi\epsilon_0 c^2 }\sum_j\int_{0 }^{ +
\infty }d\Omega_j \Omega_j \mu(\Omega_j) (1 - \frac{c^2
}{{n^2v^2}}[1 + \frac{{\hbar \Omega_j }}{{2mc^2 }}(n^2 - 1)\sqrt {1
- \frac{{v^2 }}{{c^2 }}} ]^2)F_T(\Omega_j).\nonumber\\
\end{eqnarray}
which in a nondispersive medium tends to the result of \cite{15}.
\section{Conclusion}
In this paper, we have generalized a Lagrangian introduced in \cite {11} to include the external charges. In
this formalism the medium is modelled with two independent collections of vector fields. The classical
electrodynamics in the presence of a polarizable and magnetizable medium is discussed and the susceptibility
functions of the medium are calculated in terms of the coupling functions. The energy loss of a point charged
particle per unit length emitted in the form of radiation, called Cherenkov radiation, is obtained in both zero
and finite temperature in classical regime. A fully canonical quantization of both electromagnetic field and the
dynamical variables, modelling the medium, is demonstrated. In Heisenberg picture, the constitutive equations of
the medium together with the Maxwell equations are obtained as the equations of motion of the total system. The
wave equation for the vector potential is solved. It is shown how vector potential operator in this theory can
be expressed in terms of the medium operators at an initial time. The consistency of these solutions for the
field operators are found to depend on the validity of certain velocity sum rules. It is also shown how this
scheme is related to the damping polarization and phenomenological quantization theories. The large-time limit
and quantum mechanical perturbation theory is applied to treat the finite temperature Cherenkov radiation in the
domain of the non-relativistic and relativistic quantum regimes. The total energy radiated per unit time is
calculated which is consistent with both classical and extreme relativistic limits. The approach is based on a
Lagrangian formalism and magnetic properties of the medium and the relativistic motion of the particle are
included. This model can be applied to the case of Cherenkov radiation in a nonlinear medium which is under
consideration.

\appendix*
\section{}
In this appendix we give the analogous expression for the vector potential in the quantum domain. Furthermore,
we evaluate the time-dependent coefficients for the vector potential operator. By using Eqs. (\ref{b18}) and
(\ref{b19}) the wave equation (\ref{b16}), can be written as
\begin{eqnarray} \label{1}
&& \mu _0 \varepsilon _0 \underline{{\rm \ddot A}} + \left| k \right|^2 \underline{{\rm A}} + \mu _0
\frac{\partial }{{\partial t}}\int_0^t {dt'\,\chi _e (t - t')\underline{{\rm \dot A}}({\rm k},t'){\rm  } - \mu
_0 \left| k \right|^2
\int_0^t {dt'\,\chi _m (t - t')\underline{{\rm A}}({\rm k},t'){\rm  }} }\nonumber \\
&& = \mu _0 \frac{{\partial \underline{{\rm P}}^{N\, \bot } }}{{\partial t}} + \mu _0 \iota {\rm k} \times
\underline{{\rm M}}^{N}+\mu_0\underline{{\bf J}}^\bot.
\end{eqnarray}
We find the full time dependence of the vector potential by taking the inverse Laplace transform. The inverse
Laplace transformation as defined before is a contour integration over the Bromwich contour. After transforming
to frequency variables we obtain
\begin{eqnarray} \label{2}
{\rm A}({\rm r},t)& =& \sum\limits_{\lambda  = 1}^2  \int {d^3 {\rm k}} \sqrt {\frac{\hbar }{{2(2\pi )^3
\varepsilon _0 ck}}} \left( {\eta ( k ,t)a_\lambda ({\rm k},0)e^{\iota {\rm k} \cdot {\rm r}}  + h.c.}
\right){\bf e}_\lambda ({\rm
k})\nonumber\\
& +& \frac{1}{{\varepsilon _0 }}\sum\limits_{\lambda  = 1}^2  \int {d^3 {\rm k}} \int d\omega\sqrt {\frac{\hbar
}{{2(2\pi )^3 \omega }}} \left( {\beta (\omega ,k,t)d_\lambda  ({\rm k},0)e^{\iota
{\rm k} \cdot {\rm r}}  + h.c.} \right){\bf e}_\lambda  ({\rm k})\nonumber\\
& +& \frac{\iota }{{\varepsilon _0 }}\sum\limits_{\lambda  = 1}^2 \int {d^3 {\rm k}}\int d\omega \sqrt
{\frac{{\hbar k^2 }}{{2(2\pi )^3 \omega }}} \left( {\gamma (\omega ,k ,t)b_\lambda ({\rm k},0)e^{\iota {\rm k}
\cdot {\rm r}}  + h.c.} \right){\bf s}_\lambda
({\rm k})\nonumber\\
& +& \sum_\alpha\sum\limits_{\lambda  = 1}^2\frac{q_\alpha}{2} \zeta (k,t)({\rm \dot r}_\alpha  \delta ({\rm r}
- {\rm r}_\alpha  ) + \delta ({\rm r} - {\rm r}_\alpha  ){\rm \dot r}_\alpha)\cdot {\bf e}_\lambda  ({\rm k})
{\bf e}_\lambda  ({\rm k})
\end{eqnarray}
where
\begin{eqnarray}\label{3}
\eta (k,t) &=& \frac{1}{{2\pi \iota }}\int_{ - \iota \infty }^{ +
\iota \infty } ds e^{st}{\frac{{s\tilde{\epsilon}(s)-\imath ck
}}{s^2 \tilde{\varepsilon} (s) + k^2 c^2 \tilde{\kappa}
(s)}}=\sum\limits_j [ {\mathop{Re}\nolimits} (e^{ - \iota \Omega _j
t} \frac{{v_g^j }}{{v_p^j }})-\imath{\mathop{ Im}\nolimits} (e^{ -
\iota \Omega _j t} \frac{{v_g^j }}{{c\kappa (\Omega _j )}})]\nonumber\\
\end{eqnarray}
\begin{eqnarray}\label{4}
\beta (\omega,k,t) &=& \frac{f(\omega)}{{2\pi \iota }}\int_{ - \iota
\infty }^{ + \iota \infty } ds{\frac{{s e^{st}}}{(s+\imath
\omega)(s^2 \tilde{\varepsilon} (s) + k^2 c^2 \tilde{\kappa}
(s))}}=\frac{-\imath\omega
f(\omega)e^{-\imath\omega t}}{-\omega^2\tilde{\varepsilon }(\omega) + k^2 c^2 \tilde{\kappa} (\omega)}\nonumber\\
\end{eqnarray}
\begin{eqnarray}\label{5}
\gamma (\omega,k,t) &=& \frac{g(\omega)}{{2\pi \iota }}\int_{ -
\iota \infty }^{ + \iota \infty } ds{\frac{{e^{st} }}{(s+\imath
\omega)(s^2 \tilde{\varepsilon} (s) + k^2 c^2 \tilde{\kappa}
(s))}}=\frac{g(\omega)e^{-\imath\omega t}}{-\omega^2\tilde{\varepsilon} (\omega) + k^2 c^2 \tilde{\kappa}
(\omega)}\nonumber\\
\end{eqnarray}
\begin{equation}\label{6}
\zeta (k,t) = \frac{1}{{2\pi \iota }}\int_{ - \iota \infty }^{ +
\iota \infty } ds{\frac{{e^{st} }}{s^2 \tilde{\varepsilon} (s) + k^2
c^2 \tilde{\kappa }(s)}}=\frac{1}{{kc}}\sum\limits_j {\mathop{
Im}\nolimits} (e^{ - \iota \Omega _j t} \frac{{v_g^j }}{{c{\kappa}
(\Omega _j )}}).
\end{equation}
The solution of the wave equation is the sum of a transient and a permanent part. The latter are expressed
solely in terms of the initial medium operators. Long after the initial time, vector potential operator will be
a function of the medium operators alone since it has poles on the imaginary axis in the complex $s$-plane. With
the velocity sum rules discussed in Sec. 2, one can see that the coefficient $\eta (k,t)$ equals 1 at time $t=0$
whereas the other coefficients have the initial value 0.

\begin{acknowledgments}
F. Kheirandish  and E. Amooghorban wish to thank The Office of Graduate Studies and Research Vice President of
The University of Isfahan for their support.
\end{acknowledgments}

\end{document}